\documentstyle[jiro,epsfig]{article}

\def\Journal#1#2#3#4{{#1} {\bf #2}, #3 (#4)}

\def\NPB{{\em Nucl. Phys.} B}
\def\PLB{{\em Phys. Lett.}  B}

\def\PRD{{\em Phys. Rev.} D}
\def\ZPC{{\em Z. Phys.} C}
\def\APP{{\em Acta Phys. Polonica} B}

\def\be{\begin{equation}}
\def\ee{\end{equation}}
\def\bea{\begin{eqnarray}}
\def\eea{\end{eqnarray}}

\begin{document}
\title{COMPLETE DESCRIPTION OF POLARIZATION EFFECTS IN TOP QUARK DECAYS 
INCLUDING HIGHER ORDER QCD CORRECTIONS 
\footnote{invited talk presented at the International Workshop 
          on QCD and New Physics, Hiroshima, 1997}
                      }
\author{BODO LAMPE}
\address{Department of Physics, University of Munich, 
Theresienstrasse 37, D--80333 Munich 
\\E-mail: bol@mppmu.mpg.de}
\maketitle
\abstracts{
The complete set of matrix elements for all polarization 
configurations in top quark decays is presented including 
higher order QCD corrections. The analysis is done in the framework 
of the helicity formalism. The results can be used in a 
variety of circumstances, e.g. in the experimental analysis 
of top quark production and decay at Tevatron, LHC and NLC. 
Relations to LEP1 and LEP2 physics are pointed out.}
\section{Introduction}
Since its discovery in 1995 the top quark has been an object  
of increasing interest. The production process for top quarks 
has been analyzed in various theoretical studies both 
for proton and $e^+e^-$ collisions. Early references on 
the lowest order cross section are \cite{combridge} for 
pp collsions and \cite{reiter} for $e^+e^-$ annihilation. 
Higher order corrections 
to the cross section (total cross section, $p_T$ distribution etc.) 
have been calculated by several groups, \cite{ellis,neerven} 
for pp and \cite{laermann,jadach} for $e^+e^-$. 
These total cross sections do not involve information 
on the top quark polarization. Such spin effects only come 
in, if one studies distributions of top quark decay products. 
In some cases, spin effects have been studied, i.e. the 
distribution of the top quark spin vector, in ref. 
\cite{parke} for pp and in ref. \cite{arens} for 
$e^+e^-$ collisions. These latter 
studies have not been extended to higher orders yet. However,  
an interesting step in this direction has been taken in  
ref. \cite{jadach}, where a Monte Carlo 
progam including final state spin terms has been written. 
Unfortunately, that paper is in fact 
concerned with higher order corrections to $\tau^+\tau^-$ production 
in $e^+e^-$ annihilation at lower energies and does not take into account 
axial vector couplings. More general results, including 
axial vector couplings, can be found in \cite{groote}, 
but not in the form of a Monte Carlo program. 

Nothing is known about higher order spin corrections to 
the processes which induce top quark production in 
proton collisions (light quark annihilation $q\bar q \rightarrow 
t\bar t$ and gluon--gluon fusion $q\bar q \rightarrow 
t\bar t$). In contrast to $e^+e^-$ annihilation, 
there are many higher order diagrams involved in these processes, 
so that the calculations are very difficult.   

Of course, in studying distributions of top quark decay products 
one has to take into account the higher order 
top quark decay matrix elements, too. 
The spin averaged matrix elements have been calculated 
including higher order corrections. 
For example, oneloop QCD corrections are known to decrease the 
total width of the top quark by about 8\%. If one studies 
distributions of top quark decay products with reference to 
the production process, one needs in addition the spin correlations 
in the top quark decay matrix elements.   
In lowest order, these spin correlations are known, 
but in higher order only rudimentary results exist 
\cite{czarnecki,lampe}. In those calculations, higher order corrections 
to special decay product distributions were calculated. This 
corresponds to certain linear combinations of the polarized 
matrix elements. The present article describes how {\it all} the 
polarized matrix elements can be obtained in a complete and systematic  
way in the framework of the so--called helicity formalism.  

Within the Standard Model, all couplings of the top quark to other 
particles are completely fixed by its mass and by a few quantum numbers.  
For example, the coupling of the top quark to gluons is a pure 
vector coupling with strength $g_s$, the coupling to the W--boson 
is a V--A coupling etc. The coupling to the W--boson is particularly 
interesting for this article, because it induces the decay 
$t\rightarrow W^+b$. b--quark mass terms $O(m_b/m_t)$ can 
very probably be neglected 
in higher order corrections to the decay process, because they are known  
to be small ($\sim$ 1\%) in leading order. In this approximation 
it can be shown that the b--quark is always left--handed in 
the top quark decay process, both in lowest order {\it and in higher order QCD}.  
As will be seen, this strongly reduces the number of independent 
helicity amplitudes and makes the results quite intuitive. 

\section{Helicity vs. Spin Vector Description} 
Helicity amplitudes have been considered in many applications 
of phenomenological importance in high energy physics, like 
jet production \cite{dixon}, nonstandard effects in top quark 
processes \cite{kane}, and many others. The idea is, first to 
separate a given process into simpler subprocesses and then to 
explicitly evaluate all  
the possible spin amplitudes for the subprocesses 
in special Lorentz and Dirac frames. The results can afterwards 
be put together with the help of a master formula (to be given below). 
For example, consider the lowest order helicity amplitudes for top quark 
decay in the Standard Model. They are given by 
\begin{equation}
A(h_t,h_W)={g \over \sqrt{2} } \bar u_{-1/2}(p_b)\gamma_{\mu}
{1\over 2}(1-\gamma_5) u_{h_t}(p_t) \epsilon^{\mu}_{h_W}
\label{eq1}
\end{equation}
where $g={e\over s_W}$ and 
$h_t$ and $h_W$ label the spins for the top quark and the 
W--boson, $h_t=\pm {1\over 2}$ and $h_W=0,\pm 1$. The helicity 
of the (massless) bottom quark is fixed to be $-1/2$ by the 
V--A nature of the interaction. Thus, in the Standard Model 
there are six amplitudes to be considered, and this number remains 
the same in higher order QCD (after integrating over the gluon degrees of 
freedom). The six amplitudes can be used to 
define the 36 elements of the density matrix 
\begin{equation}
\rho (h_t,h_W,h_t',h_W'):=A(h_t,h_W)A(h_t',h_W')^{\ast} \, .
\label{dm}
\end{equation} 
Note that the amplitudes are only determined up to an overall phase, 
and that 
this arbitrariness goes away when forming the density matrix elements.  
Unfortunately, higher order QCD corrections  
cannot be fully calculated on the level of amplitudes, but only 
on the level of the density matrix which contains in principle 
36 degrees of freedom. As will be discussed below, hermiticity and  
CP invariance reduce this number, 
but still leave an appreciable set of 
matrix elements. 

The full density matrix is in fact needed in the 'master formula', 
if one considers some combined production and decay process 
for the top quark. Namely, assume that top quarks are produced 
in some process $ab \rightarrow t\bar t$ and then decay 
according to $t \rightarrow W^+b$ and $\bar t \rightarrow W^-\bar b$, 
where the W's further decay to light (massless) fermions, $ W^+ \rightarrow 
f_1 \bar f_2$ and $ W^- \rightarrow 
f_3 \bar f_4$. The cross section is then given by the 'master formula' 
\begin{equation}
\sigma =\sum_{EXT} \vert \sum_{INT} A(h_a,h_b,h_t,h_{\bar t})
A(h_t,h_{W^+})A(h_{\bar t},h_{W^-})A(h_{W^+})A(h_{W^-})\vert ^2
\label{mf}
\end{equation}
where $EXT=h_a,h_b$ denotes the spins of the external 
particles and $INT=h_t,h_{\bar t},h_{W^+},h_{W^-}$ the spins of 
the internal particles of the process. 
$A(h_a,h_b,h_t,h_{\bar t})$ are the helicity amplitudes for the 
production process, $A(h_{\bar t},h_{W^-})$ for the decay of the 
antitop quark, and $A(h_{W^+})$ and $A(h_{W^-})$ are the amplitudes 
for the decay of the $W^+$ and $W^-$, respectively. Note that 
before Eq. (\ref{mf}) the $W^+$ spin has been denoted by $h_W$ 
instead of $h_{W^+}$, and for simplicity it will again be 
denoted by $h_W$ below. The fact that
the formula makes no explicit reference to the spins of the 
massless fermions $f_i$, i=1,2,3,4, has the same reason as 
the non--appearance of the b--quark spins. Namely, the $h_{f_i}$ are 
fixed by the V--A nature of the W decays. 
Furthermore, if one is not interested in the decay of the antitop 
or of the W's, the corresponding amplitudes and helicities will 
not appear in  
the above formulas. In that case, the $\bar t$ and/or the W's 
will be one of the {\it external} (EXT) particles, whose spins 
have to be summed over after taking the square in Eq. (\ref{mf}).    

Besides neglecting the b--quark mass, it is also a good 
approximation in Eq. (\ref{mf}) to take the internal particles 
on--shell, because off-shell contributions are suppressed by powers 
of the width $\Gamma_t$ and $\Gamma_W$. More precisely, one 
has 
\begin{equation}
{1\over (P^2-M^2)^2+M^2\Gamma^2}={\pi \over M\Gamma}
\delta (P^2-M^2) +O({\Gamma\over M})
\label{nw}
\end{equation}
for a particle of mass $M$ and 4--momentum $P$. For a top quark 
of mass 175 GeV, the width is about 1.5 GeV, so that terms of 
order ${\Gamma\over M}$ can be neglected, in particular in higher 
orders where $O(\alpha_s {\Gamma\over M})$ is of the order of a permille. 
Similar considerations apply to the W--boson. 

In higher order 
the narrow width approximation Eq. (\ref{nw}) is of particular use in 
reducing the number of diagrams to be calculated. The reason is that 
all diagrams where a gluon runs from the production part of the 
process to the decay part, and also the corresponding interference 
diagrams, give contributions which are suppressed by powers 
of ${\Gamma\over M}$. Therefore, in this approximation the 
process can be really decomposed into a number of building 
blocks, the production block, the t--decay blocks and the 
W--decay blocks. In the narrow width approximation, 
these blocks are interrelated by spin--indices, 
but not by gluon exchange.  

When one carries out the modulus squared in 
Eq. (\ref{mf}), it becomes apparent that in general the 
full density matrix $A(h_t,h_W)A(h_t',h_W')^{\ast}$ is needed 
to calculate the cross section of the decay products.  

I have calculated the six amplitudes $A(h_t,h_W)$ 
using the 'chiral representation' 
of $\gamma$ matrices, in which $\gamma_5=$diag$(-1,-1,1,1)$ etc. 
This representation makes calculations with massless 
fermions (the b--quark in the case at hand) quite transparent. 
Furthermore, I shall present results in two different Lorentz 
frames, in the rest frame of the top quark and the rest frame 
of the W--boson. Both frames have special virtues, so it is worthwhile 
to study them both. The top quark rest frame is of course 
the natural frame to study top decays, and to look at 
distributions in the energies of the decay products etc. 
The W rest frame has the particular virtue that the amount 
of longitudinal W's can be read off most easily in this frame. 
I am quite sure there is a (complicated) transformation between 
the amplitudes in both frames. However, I was not able to 
derive it and, furthermore, found it reasonably  
convenient to do the calculations in both frames separately. 
  
There is a popular alternative to the helicity formalism, 
where use is made by the fact that the spin of a fermion 
with 4--momentum P 
can be described by a 'spin vector' S, a pseudo 4--vector 
which fulfills $S^2=-1$ and $S\cdot P=0$.  
In this formalism one does not calculate amplitudes but 
(squared) matrixelements. The matrix element 
for the production of a $t\bar t$ pair with spin vectors 
$s_t$ and $s_{\bar t}$ has the generic form 
\begin{eqnarray} \nonumber 
\vert M \vert_P^2 & \sim & tr (p\llap{/}_t+m_t)(1+\gamma_5 s\llap{/}_t) 
         \ldots  \ldots 
                           (p\llap{/}_{\bar t}+m_t)(1+\gamma_5 s\llap{/}_{\bar t})
 \ldots  \ldots \\ 
& = & a+bs_t+cs_{\bar t}+ds_ts_{\bar t}
\label{sv1}
\end{eqnarray}
where, for example, $d$ is a tensor with two Lorentz indices and $b$ and $d$  
are 4--vectors. Similarly, 
the matrix elements for the decay of the top quarks have the 
form 
\begin{equation} 
\vert M \vert_D^2 =e+fs_t \qquad 
\vert M \vert_{\bar D}^2 =\bar e+\bar fs_{\bar t} 
\label{sv2}
\end{equation}
and the full matrix element for production and decay 
is then given by 
\begin{equation} 
\vert M \vert =ae\bar e -bf\bar e -ce\bar f +df\bar f  \, .
\label{sv3}
\end{equation}
According to this formula, the cross section will not be 
just a product of a production and of a decay piece, but 
is given by a sum of such products. In fact, Eq. (\ref{sv3}) 
can be shown to be equivalent to Eq. (\ref{mf}) \cite{bjorken}. 

\section{The Method}
The results to be presented were obtained in the helicity 
formalism. Furthermore, they concern the oneloop QCD corrections 
to the lowest order matrix elements. The lowest order expressions 
are usually simple, whereas the oneloop expressions, in particular 
for the case of hard gluons are quite lenghty for arbitrary 
spin orientations. Furthermore, the hard gluon contributions 
cannot be treated on the level of amplitudes, because they 
have to be integrated over the gluon's energy and angles. 
One really has to go to the level of the (spin) density 
matrix, Eq. (\ref{dm}), to do the phase space 
integrations. There is, however, one circumstance which simplifies 
the task. This is related to the fact, that the oneloop 
QCD corrections to the total (spin--averaged) width of the 
top quark are known \cite{dennersack}. This allows to get 
rid of the infrared and collinear singularities present 
in the matrix elements, by forming suitable singularity--free 
combinations of the spin--dependent and the spin--averaged 
expressions. The point is that the infrared and collinear 
singularities are 'universal', i.e. independent of the spin 
direction, so that they will drop out in suitable differences. 
We shall discuss our procedure in more detail in the next section, 
where QCD corrections to $W\rightarrow q\bar q^\prime $ are 
considered as a rather simple warming up exercise.   

\section{QCD Corrections to $W\rightarrow q\bar q^\prime $}
This process is simpler because both outgoing quarks can be 
considered to be massless particles, so that their helicities 
are fixed by the $V-A$ nature of the decay. Therefore, the decay amplitudes 
depend only on the W--spin and in lowest order are given by 
\begin{equation} 
A(0)=\sin \theta      \qquad  
A(\pm 1)={1\pm \cos \theta \over \sqrt{2}  }e^{\pm i\phi}     
\label{wd1}
\end{equation}
where $\theta$ and $\phi$ are the (polar and azimuthal) angles 
between the z--direction (defined as the direction 
of the outgoing quark $q$) and the direction of the 
W as given in some LAB frame (it may also 
be considered as the direction to which the W--spin points). 
These formulae hold in the rest frame of the W--boson. 
An overall factor ${em_W\over 2\sqrt{2}s_W}$ has been left out 
in the amplitudes. 
Note that the integrated W--width 
$\Gamma_W={e^2m_W^2 \over 48\pi s_W^2}$ can be obtained from 
the trace of the density matrix $\sum_{h_W}A(h_W)A(h_W)^\ast$ 
by multiplying with the square of the factor 
${em_W\over 2\sqrt{2}s_W}$ and by dividing by the well--known  
$16\pi m_W$ \cite{particledatabook}.  
 
Using Eq. (\ref{wd1}), the corresponding density matrix 
$D_{h_W,h_W'}\equiv A(h_W)A(h_W')^\ast$ can easily 
be calculated in lowest order to yield 
\begin{eqnarray} \nonumber   
D^{lo}(0,0)&=&\sin^2\theta \\ \nonumber 
D^{lo}(\pm 1,\pm 1)&=&{1 \over 2}(1+\cos^2\theta) 
                 \underline{\pm\cos\theta} \\ \nonumber 
D^{lo}(+1, -1)&=&{1 \over 2}\sin^2\theta e^{2i\phi} \\ 
D^{lo}(\pm 1, 0)&=& 
(\pm\cos\theta\sin\theta \underline{-\sin\theta}   ) 
{ e^{\pm i\phi} \over \sqrt{2}}   \,  . 
\label{wd2}
\end{eqnarray} 
The underlined terms refer to parity violating effects and 
cannot be measured in hadronic W decays,  
because the outgoing quark is detected in the form of a jet and 
its flavor and charge cannot be identified.  
The functions $D(h_W,h_W')$ are sometimes called the decay functions of the  
W. An upper index $lo$ has been introduced in Eq. (\ref{wd2}) 
in order to make clear 
that these are the Born level contributions to the density matrix. 
The aim is then to calculate higher order corrections to the 
decay functions/density matrix, in the 
form 
\begin{equation}  
D(h_W,h_W')=
D^{lo}(h_W,h_W')
+{\alpha_s \over \pi}D^{ho}(h_W,h_W') 
\label{wd3}
\end{equation}
To accomplish this calculation, use was made of the well--known 
higher order result for the spin averaged decay function, the 
'trace of the spin density matrix', which in our normalization 
is given by 
\begin{equation}  
D_{total}=D^{lo}_{total}+{\alpha_s \over \pi}D^{ho}_{total}
=2(1+{\alpha_s \over \pi})  \, . 
\label{wd4}   
\end{equation}
The point to notice is that one has 
\begin{equation}
{D(h_W,h_W')\over D_{total}}={D^{lo}(h_W,h_W')\over D^{lo}_{total}} +
{\alpha_s \over \pi} 
{D^{ho}(h_W,h_W')D^{lo}_{total}-D^{lo}(h_W,h_W')D^{ho}_{total} \over 
                      (D^{lo}_{total})^2  }
\label{wd5}               
\end{equation}
where the difference 
$D^{ho}(h_W,h_W')D^{lo}_{total}-D^{lo}(h_W,h_W')D^{ho}_{total}$ 
is completely free of singularities separately for hard and 
virtual gluons. In fact, the diagrams with virtual gluon exchange 
contribute nothing to the ratio ${D(h_W,h_W')\over D_{total}}$. 
(In top decay, where we shall proceed similarly, the virtual 
diagrams will contribute, but only a finite amount.) 
The real gluon processes $W\rightarrow q\bar q' g$ can be explicitly 
seen to give a finite contribution to the above difference, i.e. 
the result is finite for $E_g\rightarrow 0$ and $\theta_g\rightarrow 0$, 
where $E_g$ and $\theta_g$ are the gluon energy and angle with 
respect to the quark direction \cite{abraham}. The integration 
over $E_g$ and $\theta_g$ is therefore straightforward and one 
obtains the corrections to the W decay functions in a very 
compact form \cite{abraham}  
\begin{equation}                         
D(h_W,h_W')=(1+{\alpha_s \over \pi})(1-0.975{\alpha_s \over \pi}) 
[D^{lo}(h_W,h_W')+0.653{\alpha_s \over \pi}\delta_{h_W,h_W'}]   \, . 
\label{wd6}              
\end{equation}
This representation can only be obtained after neglecting the irrelevant 
parity violating terms. Furthermore, 
in higher 
orders the angles $\theta$ and $\phi$ have been defined as to 
refer to the thrust--  
instead of the quark--momentum direction.

These results have been applied to W--pair production at 
LEP2 \cite{abraham} and NLC \cite{abraham1}. 
According to a master formula similar to (\ref{mf}),   
the cross section for W--pair production and decay in $e^+e^-$ 
annihilation was calculated including the higher order corrections 
Eq. (\ref{wd6}) and nonstandard contributions. 
Our main motivation in studying this cross section 
was twofold:  
\begin{itemize}
\item
first of all we wanted to know how QCD corrections to 
angular correlations of W decay products differ from the 
naive expection of a constant K--factor $\sim 1+{\alpha_s \over \pi}$.  
We found that depending on the kinematic point, the deviations 
from a constant K--factor can be appreciable (of the order of 
a few percent). Unfortunately, at LEP2 with its few thousand 
W events these effects are just at the edge to become visible. 
The situation is different at NLC with its larger statistics, 
where the QCD corrections really become relevant.  
\item 
secondly we have proven that QCD correction can mimic the 
presence of nonstandard physics. As has been shown by 
Monte Carlo studies \cite{algueriz}, NLC is sensitive to 
nonstandard couplings as small as $10^{-3}$. This is well 
below the magnitude of the QCD effects induced by Eq. (\ref{wd6}) 
\cite{abraham1}.  
\end{itemize}

In addition to the QCD corrections presented above, there 
are off-shell W corrections (with and without QCD \cite{pittau}) 
which are particularly important at LEP2, i.e. near threshold. 
Unfortunately, there is no space here to discuss them in detail. 

\section{Complete Lowest Order Analysis 
of the Spin Density Matrix for $t\rightarrow bW$}

The decay of the top quark has some extremely interesting 
physics features, which are related to the spin decomposition 
of the matrix element. In particular, it is well known  
that, due to the large top quark mass value, 
t--decay is dominated by longitudinal W's. 
In fact, the total width of the W is given by 
\begin{equation}                 
\Gamma_t=\Gamma_L+\Gamma_T={G_{F} m_{t}^{3} \over 8 \sqrt{2} \pi }
          (1-{m_W^2\over m_t^2})^2
          (1+2{m_W^2\over m_t^2}) 
\label{q1}                      
\end{equation}            
and the ratio of the number of longitudinal over transverse 
W's is given by $\Gamma_L/\Gamma_T={{m_t^2} \over {2m_W^2}}$. 
Note that the QCD corrections to $\Gamma_t$, $\Gamma_L$ and 
$\Gamma_T$ are known \cite{dennersack,lampe} to be about 
-9\% resp. 5\%, and that all other correction effects like 
electroweak, b--mass and finite width effects contribute roughly 
1\%. Note further that $\Gamma_L$ is the 
most interesting source of loop corrections to the famous  
$R_b$ value [the partial width $\Gamma (Z\rightarrow b\bar b)$ 
measured at LEP1]. 
The point is that the exchange of longitudinal W's between 
the b-quarks gives rise to the celebrated corrections 
of order $O(G_Fm_t^2)\sim {{m_t^2} \over {m_W^2}}$.  
Unfortunately, at LEP1 
this is only a small loop effect which is of the order 
of 1\%, because it is 
suppressed by a factor 
${1\over 16\pi^2}$. In contrast, in t--decays the longitudinal 
W's enter as the most dominant leading order effect, so that 
they can be studied much clearer.  

Top quark decay may be looked at in different frames; particularly 
interesting are the rest frames of the t--quark or that of the 
W. In the W rest frame, for example, there is a very simple 
way to experimentally determine the ratio $\Gamma_L/\Gamma_T$ 
and other spin--dependent observables \cite{lampe}.              
Of course, the two systems are related just by a simple boost. 
However, the transformation formula between the spin--amplitudes 
in the two systems is quite complicated, so that I prefer to give 
results in the two systems separately. 
Let's start with the t rest system.  
In order to calculate the amplitudes (\ref{eq1}), 
I have choosen the following parametrization of momenta, polarization 
vectors and spinors. First, the momenta:   
\begin{equation}
p_t=(m_t,\vec 0) \qquad p_W={m_t\over 2}(f_+,0,0,f_-) 
\qquad p_b=p_t-p_W 
\label{q2}
\end{equation}
where $f_{\pm}=1\pm f$ and $f={{m_W^2} \over {m_t^2}}$ and 
the W--boson has been chosen to define the z--direction. The most 
general top quark spinor is given by 
\begin{eqnarray} \nonumber
u_{+1/2}(p_t)&=&\sqrt{m_t}
          (\cos {\theta \over 2}, \sin {\theta \over 2}e^{i\phi},
          \cos {\theta \over 2}, \sin {\theta \over 2}e^{i\phi}) 
\\ 
u_{-1/2}(p_t)&=&\sqrt{m_t}
         (-\sin {\theta \over 2}e^{-i\phi}, \cos {\theta \over 2}, 
          -\sin {\theta \over 2}e^{-i\phi}, \cos {\theta \over 2})  
\label{q3}
\end{eqnarray}  
where $\theta$ and $\phi$ refer to some direction, e.g. to the  
direction of the top quark in some lab--system. 
One may put $\phi =0$ without restriction, because this 
corresponds to defining the y--direction. 
The possible b-- and W--polarizations are fixed to be 
\begin{equation}
\bar u_{-1/2}(p_b)=\sqrt{m_t}(0,0,\sqrt{f_-},0)
\label{q4}
\end{equation}
and 
\begin{equation}
\epsilon_{\mp 1}=-{1\over \sqrt{2}}(0,\pm 1,i,0) 
\qquad \epsilon_0=-{1\over 2\sqrt{f}}(f_-,0,0,f_+) \, . 
\label{q5}
\end{equation}
This leads to the following amplitudes for t--decays in the 
top quark rest frame: 
\begin{eqnarray} \nonumber
&A_t^{lo}(-{1\over 2},0)= {1\over \sqrt{f}}\sin {\theta \over 2}e^{-i\phi} 
\qquad 
&A_t^{lo}(+{1\over 2},0)=- {1\over \sqrt{f}}\cos {\theta \over 2} 
\\ \nonumber 
&A_t^{lo}(-{1\over 2},+1)=0 \qquad &A_t^{lo}(+{1\over 2},+1)=0 
\\  
&A_t^{lo}(-{1\over 2},-1)=- \sqrt{2}\cos {\theta \over 2} 
\qquad 
&A_t^{lo}(+{1\over 2},-1)=- \sqrt{2}\sin {\theta \over 2}e^{i\phi}
\label{q6}
\end{eqnarray}  
where the upper index  refers to 'lowest order' and the 
lower index to top quark decay in the t rest frame.  
A universal spin independent coefficient $c_0={em_t\over 2s_W}\sqrt{f_-}$ 
has been left out in all the amplitudes. 
Note that the amplitudes are only determined up to an overall phase,  
and that 
this arbitrariness goes away when forming the density matrix elements.     
Since the amplitudes are explicitly given, it is straightforward 
to obtain the density matrix in lowest order. Its trace 
is easily obtained from the above expressions to be 
\begin{equation}                                  
\sum_{h_t,h_W}A_t^{lo}(h_t,h_W)A_t^{lo}(h_t,h_W)^\ast
= c_0^2 {1+2f\over f}   
\label{q7}                                        
\end{equation}       
and one can reproduce from this  
the total width of the top quark, Eq. (\ref{q1}) 
by dividing by the phase space factor ${f_-\over 16\pi m_t}$ 
\cite{particledatabook}.  
  
There are lots of other combinations of density 
matrix elements to describe interesting physics. 
For example, the above mentioned ratio ${\Gamma_L\over \Gamma_T}$ 
is obtained as 
\begin{equation} 
{\Gamma_L\over \Gamma_T}={          
\sum_{h_t=\pm {1\over 2},h_W=0}\vert A_t^{lo}(h_t,h_W)\vert ^2
\over 
\sum_{h_t=\pm {1\over 2},h_W=\pm 1}\vert A_t^{lo}(h_t,h_W)\vert ^2
}
={1\over 2f}  \, . 
\label{q8}
\end{equation}

Let us now repeat the same analysis in the W rest frame. 
This time I chose to define the z--direction 
by the direction of the top quark, i.e. 
\begin{equation}
p_W=(m_W,\vec 0) \qquad p_t={m_t\over 2\sqrt{f}}(f_+,0,0,f_-)
\qquad p_b=p_t-p_W \, . 
\label{q21}
\end{equation}
The top and bottom quark spinors are then fixed as 
\begin{equation}
u_{+1/2}(p_t)=\sqrt{m_t} (f^{1/4},0,f^{-1/4},0) 
\qquad 
u_{-1/2}(p_t)=\sqrt{m_t} (0,f^{-1/4},0,f^{1/4})   
\label{q217}
\end{equation}
and 
\begin{equation}
\bar u_{-1/2}(p_b)=\sqrt{m_t} (0,0,0,\sqrt{ {f_-\over \sqrt{f} }}) 
\label{q22}
\end{equation}
whereas the W polarization direction is arbitrary: 
\begin{eqnarray} \nonumber
\epsilon_{- 1}&=&-{e^{i\phi}\over \sqrt{2}}(0,
\cos\phi\cos\theta -i\sin\phi,\sin\phi\cos\theta+i\cos\phi,-\sin\theta) 
\\ \nonumber 
\epsilon_{+1}&=&-\epsilon_{- 1}^\ast 
\\ 
\epsilon_0&=&-(0,\sin\theta\cos\phi,\sin\theta\sin\phi,\cos\theta) \, . 
\label{q52}
\end{eqnarray}
The angles $\theta$ and $\phi$ refer to some arbitrary 
direction, e.g. to the
direction of the W--boson in some lab--system. Note that 
although I am using the same symbols $\theta$ and $\phi$ as in 
Eqs. (\ref{q6}) and (\ref{wd1}), 
the meaning of these angles is completely 
different in the three cases. As before,  
one may in principle put $\phi =0$ without restriction, because this
corresponds to defining the y--direction.
One is lead to the following amplitudes for t--decays in the 
W rest frame: 
\begin{eqnarray} \nonumber
A_W^{lo}(-{1\over 2},0)&=&{1 \over \sqrt{f}} \cos\theta 
\qquad 
A_W^{lo}(+{1\over 2},0)=- \sin\theta e^{i\phi} 
\\ \nonumber 
A_W^{lo}(-{1\over 2},+1)&=&{1 \over \sqrt{2f}}\sin\theta e^{-i\phi} 
\qquad 
A_W^{lo}(+{1\over 2},+1)={1 \over \sqrt{2}} (1+\cos\theta )
\\  
A_W^{lo}(-{1\over 2},-1)&=&-{1 \over \sqrt{2f}}\sin\theta e^{i\phi}
\qquad 
A_W^{lo}(+{1\over 2},-1)={1 \over \sqrt{2}} (1-\cos\theta )e^{2i\phi}
\label{q66}
\end{eqnarray}  
Again, the 
universal coefficient $c_0={em_t\over 2s_W}\sqrt{f_-}$ has been 
left out in all the amplitudes.
From the trace of the corresponding density matrix one can 
reconstruct the total width of the top quark, just as in (\ref{q7}). 
However, the ratio (\ref{q7}) can only 
be obtained for $\theta =0$, because otherwise the notion of 
'longitudinal' does not refer to the heavy quark direction.

\section{QCD Corrections to the Spin Density Matrix of $t\rightarrow bW$ 
from the Exchange of Virtual Gluons}

The oneloop QCD Corrections to t--decay are somewhat more complicated 
than to W--decay because even neglecting the b--mass 
there is one more mass parameter involved. 
This is despite the fact that the Feynman diagrams needed 
are exactly the same as for W--decay (with the directions of 
the W and one of the quarks interchanged). Namely, 
there are the 'virtual gluon' vertex and self--energy diagrams 
and two 'real' diagrams with gluon emission from one of the 
quark legs. As discussed in the introduction, there are 36 
density matrix elements (\ref{dm}) which will be considered in the normalized 
form  
\begin{equation} 
\rho_{norm} (h_t,h_W,h_t',h_W')={\rho (h_t,h_W,h_t',h_W')\over 
\rho_{total}}
\label{dm8}
\end{equation} 
because this helps to cancel universal contributions, 
in analogy to the case of W decay, Eq. (\ref{wd5}). 
$\rho_{total}\equiv \sum_{h_t,h_W}\vert \rho (h_t,h_W,h_t,h_W) \vert $ 
is defined to be the trace of the density matrix and related 
to the total width $\Gamma_t$ of the top quark as discussed above.   

Let us start the discussion with the virtual  
contributions, because they can easily be obtained from 
the corrections to the V--A vertex calculated a long time ago 
\cite{finkelstein} 
\begin{equation} 
\Gamma_\mu (t\rightarrow bW)=-{ie\over \sqrt{2}s_W} 
\bigl\{ H\gamma_\mu {1\over 2}(1-\gamma_5)+\alpha_d H_+ 
           {i\sigma_{\mu \nu}p_W^\nu \over 2m_t} {1\over 2}(1+\gamma_5)
\bigr\} 
\label{vrt}
\end{equation} 
where the known function 
$H=1+O(\alpha_s)$ 'renormalizes' the V--A structure, and 
contains all the ultraviolet, infrared and collinear 
singularities, and 
\begin{equation}
H_+=-{\ln f_- \over f} 
\label{vrt1}
\end{equation}
is a regular function 
of $f=m_W^2/m_t^2$. 
Note that the appropriate expansion parameter in the 
case of top quark decay is $\alpha_d =-C_F{\alpha_s \over 2\pi}$.   
It should further be noted that the contribution of    
$H$ to the normalized density matrix $\rho_{norm}$ vanishes because 
it cancels between the numerator and denominator in Eq. (\ref{dm8}). 
The argument works in the same way as was discussed in the case 
of the W--boson. The only difference is that now a finite  
contribution $\sim H_+$ from the $\sigma_{\mu \nu}$ term survives due 
the nonvanishing (top--)quark mass. 
In turn, one may conclude that the contribution from real 
gluon emission to the normalized density matrix is finite, too, 
because of the Lee Nauenberg theorem, which says that any such 
singularity cancels between real and virtual corrections. 
One may write down the contribution $\sim H_+$ to the amplitudes 
in the form 
\begin{equation}
A_t(h_t,h_W)=A_t^{lo}(h_t,h_W)+\alpha_d A_t^+(h_t,h_W)H_+  
\label{vrt2}
\end{equation}
where the lowest order amplitudes $A_t^{lo}$ were given in (\ref{q6}) and the 
higher order amplitudes $A_t^{+}$ originating from the $\sigma_{\mu \nu}$ term 
are given by  
\begin{eqnarray} \nonumber
&A_t^{+}(-{1\over 2},0)=- {\sqrt{f}\over 2}\sin {\theta \over 2}e^{-i\phi}
\qquad
&A_t^{+}(+{1\over 2},0)={\sqrt{f}\over 2}\cos {\theta \over 2}
\\ \nonumber
&A_t^{+}(-{1\over 2},+1)=0 \qquad &A_t^{+}(+{1\over 2},+1)=0
\\ 
&A_t^{+}(-{1\over 2},-1)={1\over \sqrt{2}}\cos {\theta \over 2}
\qquad
&A_t^{+}(+{1\over 2},-1)={1\over \sqrt{2}}\sin {\theta \over 2}e^{i\phi}
\label{q6s}
\end{eqnarray}
Again, the 
universal coefficient $c_0={em_t\over 2s_W}\sqrt{f_-}$ has been 
left out in all the amplitudes.
These results imply a   
higher order contribution due to the $\sigma_{\mu \nu}$ term 
on the level of the density matrix $\rho (h_t,h_W,h_t',h_W')$. 
Including the lowest order piece it 
will be 
of the form 
\begin{eqnarray} \nonumber 
A_t^{lo}(h_t,h_W)A_t^{lo}(h_t',h_W')^\ast +
\alpha_d H_+\bigl[A_t^{lo}(h_t,h_W)A_t^{+}(h_t',h_W')^\ast
\\
                        +A_t^{+}(h_t,h_W)A_t^{lo}(h_t',h_W')^\ast \bigr] 
\label{vrt12}
\end{eqnarray}
Note that to calculate $\rho_{norm}$, one has to divide afterwards by 
$\rho_{total}$, whose contribution from the $\sigma_{\mu \nu}$ term 
(including lo) can be calculated to be  
\begin{eqnarray} \nonumber 
\sum_{h_t,h_W}\vert A_t^{lo}(h_t,h_W)\vert ^2 
+{\alpha_s \over 2\pi}H_+\sum_{h_t,h_W} 
\bigl[A_t^{lo}(h_t,h_W)A_t^{+}(h_t,h_W)^\ast  
\\
                        +A_t^{+}(h_t,h_W)A_t^{lo}(h_t,h_W)^\ast \bigr]  
=c_0^2 \bigl\{ {1+2f\over f} -3\alpha_d H_+ \bigr\}  
\label{vrt18}
\end{eqnarray} 
This result was obtained by summing over all spin configurations 
and using the explicit representation of the amplitudes given above. 
It shows explicitly that the contribution from virtual 
gluon exchange to $\rho_{norm}$ is completely under control. 

The amplitudes corresponding to the $\sigma_{\mu \nu}$ term in the 
vertex may also be evaluated in the W--rest frame 
\begin{eqnarray} 
&A_W^{+}(-{1\over 2},0)=- {\sqrt{f}\over 2}\cos \theta
\qquad
&A_W^{+}(+{1\over 2},0)={1\over 2}\sin \theta e^{i\phi}
\\ \nonumber
&A_W^{+}(-{1\over 2},+1)=-{\sqrt{f}\over 2\sqrt{2}} \sin \theta e^{-i\phi}
\qquad 
&A_W^{+}(+{1\over 2},+1)=-{1\over 2\sqrt{2}}(1+\cos \theta )
\\ \nonumber
&A_W^{+}(-{1\over 2},-1)={\sqrt{f}\over 2\sqrt{2}}\sin \theta e^{i\phi}
\qquad
&A_W^{+}(+{1\over 2},-1)=-{1\over 2\sqrt{2}}(1-\cos \theta )e^{2i\phi}
\label{q6sss}
\end{eqnarray}
and analgous relations as (\ref{vrt12}) and (\ref{vrt18}) apply. 
As discussed before, the angles $\theta$ and $\phi$ have 
completely different meanings in the $t$ and $W$ rest frames. 

\section{QCD Corrections to the Spin Density Matrix of $t\rightarrow bW$  
from Real Gluon Emission}
The amplitudes and density matrix for real gluon emission are the 
most difficult part of the higher order calculation and are also 
the most difficult to document. The point is not only that the amplitudes 
are quite complicated expressions, but also that one has to 
deal with the phase space integration over the real gluon's degrees  
of freedom on the level of the spin density matrix. 
Furthermore, the cancellations of the singularities in 
$\rho_{norm}$ in Eq. (\ref{dm8}) require a subtle understanding 
of the interplay between lowest order and first order QCD. 
All this enforces the 
use of an algebraic computer program like FORM, REDUCE or 
MATHEMATICA to handle the long and complicated expressions. 
After calculating the spin density matrix, the real gluon's degrees
of freedom have to be integrated over. 
This integration can be done without regularization, because 
according to the last section the correction to $\rho_{norm}$ from 
real gluons is finite and of the generic form, cf. Eq. (\ref{wd5}),  
\begin{equation}
\rho_{norm}(h_t,h_W,h_t',h_W')=\rho_{norm}^{lo}+ 
\alpha_d  
{\rho^{ho}\rho^{lo}_{total}-\rho^{lo}\rho^{ho}_{total} \over
                      (\rho^{lo}_{total})^2  }
\label{wd51}               
\end{equation}
where $\alpha_d =-C_F{\alpha_s \over 2\pi}$.
The dependence on the gluon spin is not considered, because it cannot 
be determined experimentally. Accordingly, the gluon 
polarizations have been summed in the standard fashion. 
 
To be more explicit, let us write the 
4--momenta relevant for the process $t\rightarrow Wbg$ 
in the rest system of the top quark as 
\begin{eqnarray} \nonumber & &   
p_t=(m_t,\vec 0) \qquad p_W={m_t\over 2}f_+x_W(1,0,0,\beta_W) 
\qquad p_b=p_t-p_W-p_g 
\\ & & 
p_g={m_t\over 2}f_-x_g
(1,\sin\theta_g\cos\phi_g,\sin\theta_g\sin\phi_g,\cos \theta_g) 
\label{real1}
\end{eqnarray}
where 
\begin{equation}
\beta_W^2=1-{4f\over f_+^2x_W^2}  
\label{wd52}               
\end{equation}
and where $\theta_g$
is given in terms of the other variables according to 
\begin{equation}
f_-x_g-f_+(1-x_W)={1\over 2}f_+x_Wf_-x_g(1-\beta_W\cos \theta_g)
\label{wd53}               
\end{equation}
Note that the meaning of $f$, $f_-$, $f_+$ etc is as in lowest order. 
The integrations over the gluon's degrees of freedom then have to 
be done with the phase space 
\begin{equation}
dPS_3(t\rightarrow Wbg)={m_t^2\pi^2\over 4}f_+f_-\int_0^1dx_g
\int_0^{{f_-^2x_g(1-x_g)\over f_+(1-f_-x_g)}} d(1-x_W)
\int_0^{2\pi} {d\phi_g \over 2\pi} 
\label{wd54}
\end{equation}

The top quark spinors may be taken as in Eq. (\ref{q3}),  
with $\phi =0$ because the y--z plane has not yet been specified. 
However, the W polarization vectors are different from the 
lowest order expressions, Eq. (\ref{q5}), because the 
W--momentum has changed. More precisely, 
the transverse polarization vectors are left unchanged, but the 
longitudinal polarization vector now reads 
\begin{equation}
\epsilon_0=-{1\over 2\sqrt{f}}(f_+x_W\beta_W,0,0,f_+x_W) 
\label{wd55}
\end{equation}
The $h_b={-1/2}$ spinor for the b--quark changes, too, but for the 
density matrix one needs only the combination 
$u_{-1/2}\bar u_{-1/2}={1\over 2}(1-\gamma_5) b\llap{/}$.  

I have calculated all the $6\times 6=36$ spin density matrix elements using 
these parametrizations 
\begin{eqnarray} & &   
(1+2f)\rho_{norm}(-{1\over2},0,-{1\over2},0)=c_- \, (1+0.188 \, \alpha_d) 
                                          -0.0246 \, c_+ \, \alpha_d 
\label{real2a}
\\ & & 
(1+2f)\rho_{norm}(+{1\over2},0,+{1\over2},0)=c_+ \, (1+0.188 \, \alpha_d) 
                                          -0.0246 \, c_- \, \alpha_d
\\ & & 
(1+2f)\rho_{norm}(-{1\over2},0,+{1\over2},0)
           =-  s_0 \,  (1+0.212  \, \alpha_d)
\\ & &
(1+2f)\rho_{norm}(+{1\over2},0,-{1\over2},0)
           = - s_0 \,  (1+0.212  \, \alpha_d)
\\ & & 
(1+2f)\rho_{norm}(-{1\over2},-1,-{1\over2},-1)=2fc_+ \, (1-0.236 \, \alpha_d)
                                          -0.00751 \, c_- \, \alpha_d
\label{ex33}
\\ & &
(1+2f)\rho_{norm}(+{1\over2},-1,+{1\over2},-1)=2fc_- \, (1-0.236 \, \alpha_d)
                                          -0.00751 \, c_+ \, \alpha_d
\\ & & 
(1+2f)\rho_{norm}(-{1\over2},-1,+{1\over2},-1)= 
2f \, s_0  \,  (1-0.216 \, \alpha_d) 
\\ & &   
(1+2f)\rho_{norm}(+{1\over2},-1,-{1\over2},-1)=                           
2f  \, s_0  \,  (1-0.216 \, \alpha_d)
%
%
\\  & & 
(1+2f)\rho_{norm}(-{1\over2},+1,-{1\over2},+1)=
        -0.00587 \, c_+ \, \alpha_d -0.0518 \, c_- \, \alpha_d
\label{rek1}
\\ & & 
(1+2f)\rho_{norm}(+{1\over2},+1,+{1\over2},+1)=
        -0.00587 \, c_- \, \alpha_d -0.0518 \, c_+ \, \alpha_d
\\ & & 
(1+2f)\rho_{norm}(-{1\over2},+1,+{1\over2},+1)=0.0460 \,  s_0 \, \alpha_d 
\\ & &
(1+2f)\rho_{norm}(+{1\over2},+1,-{1\over2},+1)=0.0460 \,  s_0 \, \alpha_d
%
%
\\ & & 
(1+2f)\rho_{norm}(-{1\over2},0,-{1\over2},+1)=-0.0150 \, s_0 \, \alpha_d
\\ & &
(1+2f)\rho_{norm}(-{1\over2},+1,-{1\over2},0)=-0.0150 \, s_0 \, \alpha_d
\\ & & 
(1+2f)\rho_{norm}(+{1\over2},0,+{1\over2},+1)=+0.0150 \, s_0 \, \alpha_d 
\\ & & 
(1+2f)\rho_{norm}(+{1\over2},+1,+{1\over2},0)=+0.0150 \, s_0 \, \alpha_d
\\ & & 
(1+2f)\rho_{norm}(-{1\over2},0,+{1\over2},+1)=0.0150 \, c_+ \, \alpha_d 
\\ & &
(1+2f)\rho_{norm}(+{1\over2},+1,-{1\over2},0)=0.0150 \, c_+ \, \alpha_d
\\ & & 
(1+2f)\rho_{norm}(+{1\over2},0,-{1\over2},+1)=-0.0150 \, c_- \, \alpha_d           
\\ & &
(1+2f)\rho_{norm}(-{1\over2},+1,+{1\over2},0)=-0.0150 \, c_- \, \alpha_d
\label{rek2}
%
%
\\ & & 
(1+2f)\rho_{norm}(-{1\over2},0,-{1\over2},-1)=- s_0 \, 
          \sqrt{2f}  \, (1+0.00600 \, \alpha_d )
\\ & &
(1+2f)\rho_{norm}(-{1\over2},-1,-{1\over2},0)=- s_0 \,                  
          \sqrt{2f}  \, (1+0.00600 \, \alpha_d )
\\ & & 
(1+2f)\rho_{norm}(+{1\over2},0,+{1\over2},-1)=+ s_0 \, 
          \sqrt{2f}  \, (1+0.00600 \, \alpha_d )
\\ & &
(1+2f)\rho_{norm}(+{1\over2},-1,+{1\over2},0)=+ s_0 \,
          \sqrt{2f}  \, (1+0.00600 \, \alpha_d )
\\ & & 
(1+2f)\rho_{norm}(-{1\over2},0,+{1\over2},-1)=-c_- \, 
          \sqrt{2f}  \, (1+0.00600 \, \alpha_d )
\\ & &
(1+2f)\rho_{norm}(+{1\over2},-1,-{1\over2},0)=-c_- \, 
          \sqrt{2f}  \, (1+0.00600 \, \alpha_d )
\\ & & 
(1+2f)\rho_{norm}(+{1\over2},0,-{1\over2},-1)=+c_+ \, 
          \sqrt{2f}  \, (1+0.00600 \, \alpha_d )
\\ & &
(1+2f)\rho_{norm}(-{1\over2},-1,+{1\over2},0)=+c_+ \, 
          \sqrt{2f}  \, (1+0.00600 \, \alpha_d )
%
%
\\ & &   
(1+2f)\rho_{norm}(-{1\over2},+1,-{1\over2},-1)\equiv  0                         
\label{rek3}
\\ & &             
(1+2f)\rho_{norm}(-{1\over2},-1,-{1\over2},+1)\equiv  0  
\\ & &                                                              
(1+2f)\rho_{norm}(+{1\over2},+1,+{1\over2},-1)\equiv  0                     
\\ & &
(1+2f)\rho_{norm}(+{1\over2},-1,+{1\over2},+1)\equiv  0  
\\ & &                                                              
(1+2f)\rho_{norm}(+{1\over2},+1,-{1\over2},-1)\equiv  0                    
\\ & &
(1+2f)\rho_{norm}(-{1\over2},-1,+{1\over2},+1)\equiv  0  
\\ & &                                                              
(1+2f)\rho_{norm}(-{1\over2},+1,+{1\over2},-1)\equiv  0
\\ & &
(1+2f)\rho_{norm}(+{1\over2},-1,-{1\over2},+1)\equiv  0  
\label{real2}              
\end{eqnarray}             
where $c_{\pm}={1\over 2}(1\pm \cos\theta)$ and 
$s_0={1\over 2}\sin\theta$ and 
where the numerical coefficients in order 
$\alpha_d =-C_F{\alpha_s \over 2\pi}$ 
have been obtained with $m_t=175$ GeV, for which $f=0.21$. 
With the help of my intergation programs I have shown that 
the $m_t$ dependence of these coefficients is in all cases moderate. 
For example, the coefficient $0.188$ in Eq. (\ref{real2a}) 
depends on ${m_t \over m_W}={1 \over \sqrt{f}  }$ 
in the way depicted in Figure 1. 
The $m_t$ dependence of the other independent QCD coefficients 
(denoted by 
$-0.0246$, $0.0212$, $-0.0150$, $0.00600$, $-0.0518$, $-0.00587$, 
$0.0460$, $-0.216$, $-0.236$ and $-0.00751$, i.e. by    
their values at $m_t=175$ GeV) are given in the figures that follow. 
Note that only lowest order and real gluons are incorporated in Eqs. 
(\ref{real2a})--(\ref{real2}).  
The virtual corrections Eqs. (\ref{vrt12}) and (\ref{q6s}) 
from the $\sigma_{\mu \nu}$ term have to be added. 
The factors $1+2f$ appearing on the left hand side are a 
relic of the fact that I am presenting the density matrix 
'normalized' to the total width/trace.  
There are several 
possibilities to make checks on this list. For example, 
I have checked that 
\begin{equation}
\sum_{h_t,h_W} \rho_{norm}(h_t,h_W,h_t,h_W)\equiv 1 
\label{wd57}
\end{equation}
is true including the oneloop QCD corrections. 
Furthermore, I have also checked that the ratio 
\begin{equation}                            
{\sum_{h_t} \rho_{norm}(h_t,0,h_t,0) \over 
\sum_{h_t} [\rho_{norm}(h_t,+1,h_t,+1)+\rho_{norm}(h_t,-1,h_t,-1)]}  
={\Gamma_L \over \Gamma_T}={1\over 2f}(1+\alpha_s \cdots)  
\label{wd58}                                
\end{equation}                              
reproduces the ratio of longitudinal over transverse W's 
as calculated in \cite{lampe} {\it including} higher order QCD corrections. 
Finally, there is the check as to the hermiticity of the density matrix,   
$\rho_{norm}(h_t,h_W,h_t',h_W')=\rho_{norm}(h_t',h_W',h_t,h_W)^\ast$. 
Note that the density matrix is real in the present case, 
because in the considered frame there is not azimuthal dependence. 

\begin{figure}
\begin{center}  
\epsfig{file=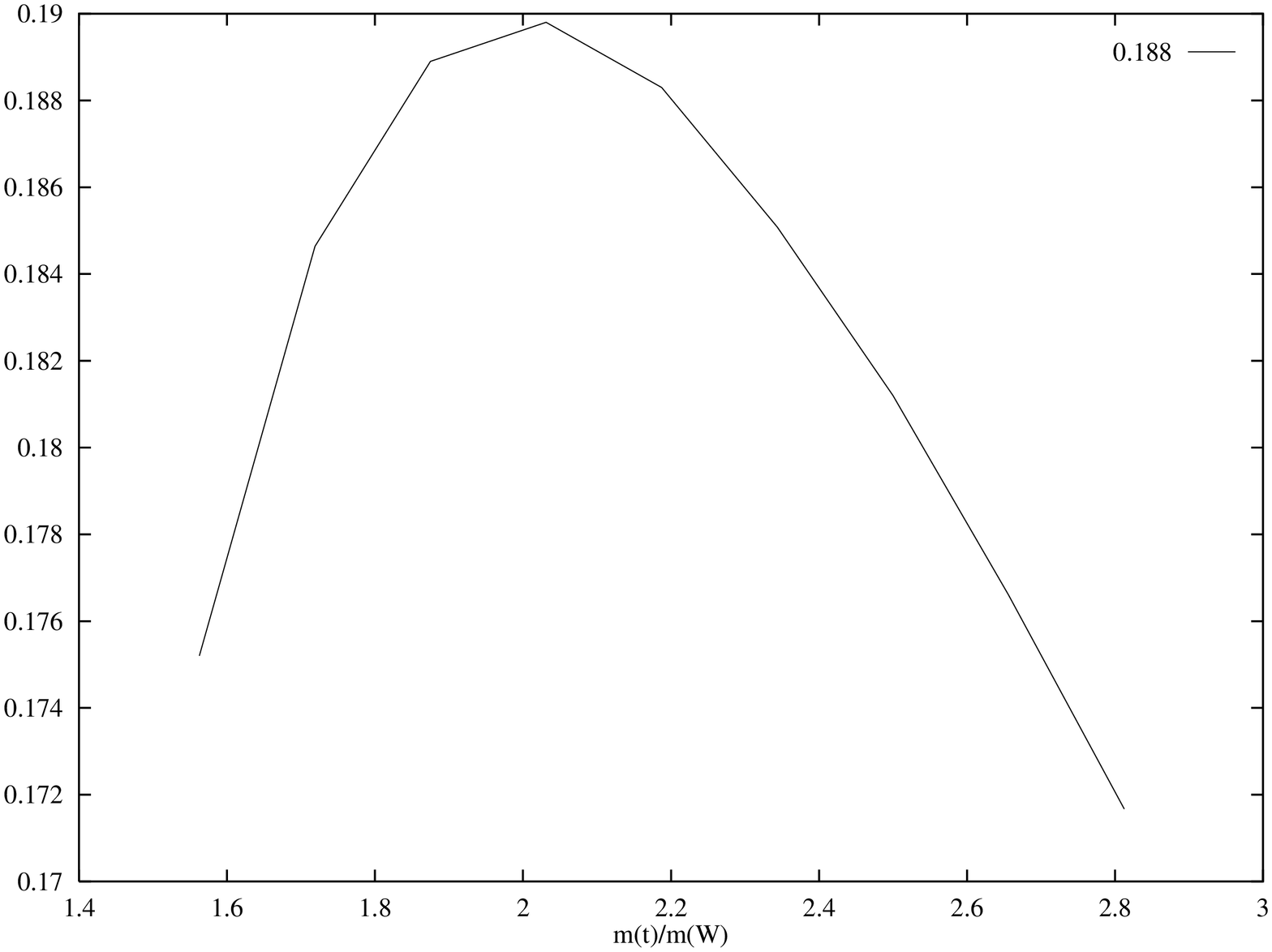,height=10cm,angle=270}
\bigskip
\caption{}
\nonumber
\end{center} 
\end{figure} 

\begin{figure}
\begin{center}  
\epsfig{file=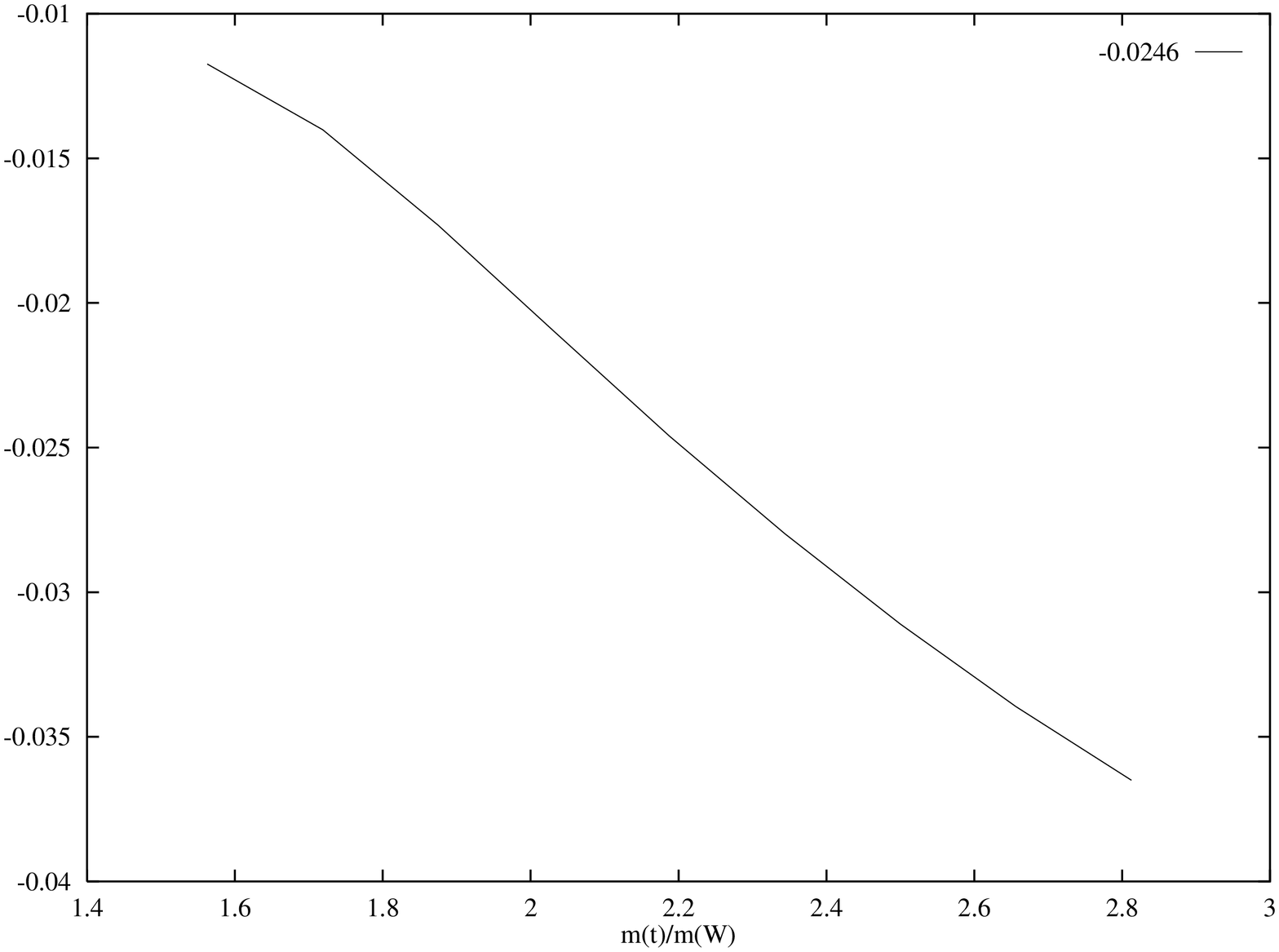,height=10cm,angle=270}
\bigskip
\caption{}
\nonumber
\end{center} 
\end{figure} 

\begin{figure}
\begin{center}  
\epsfig{file=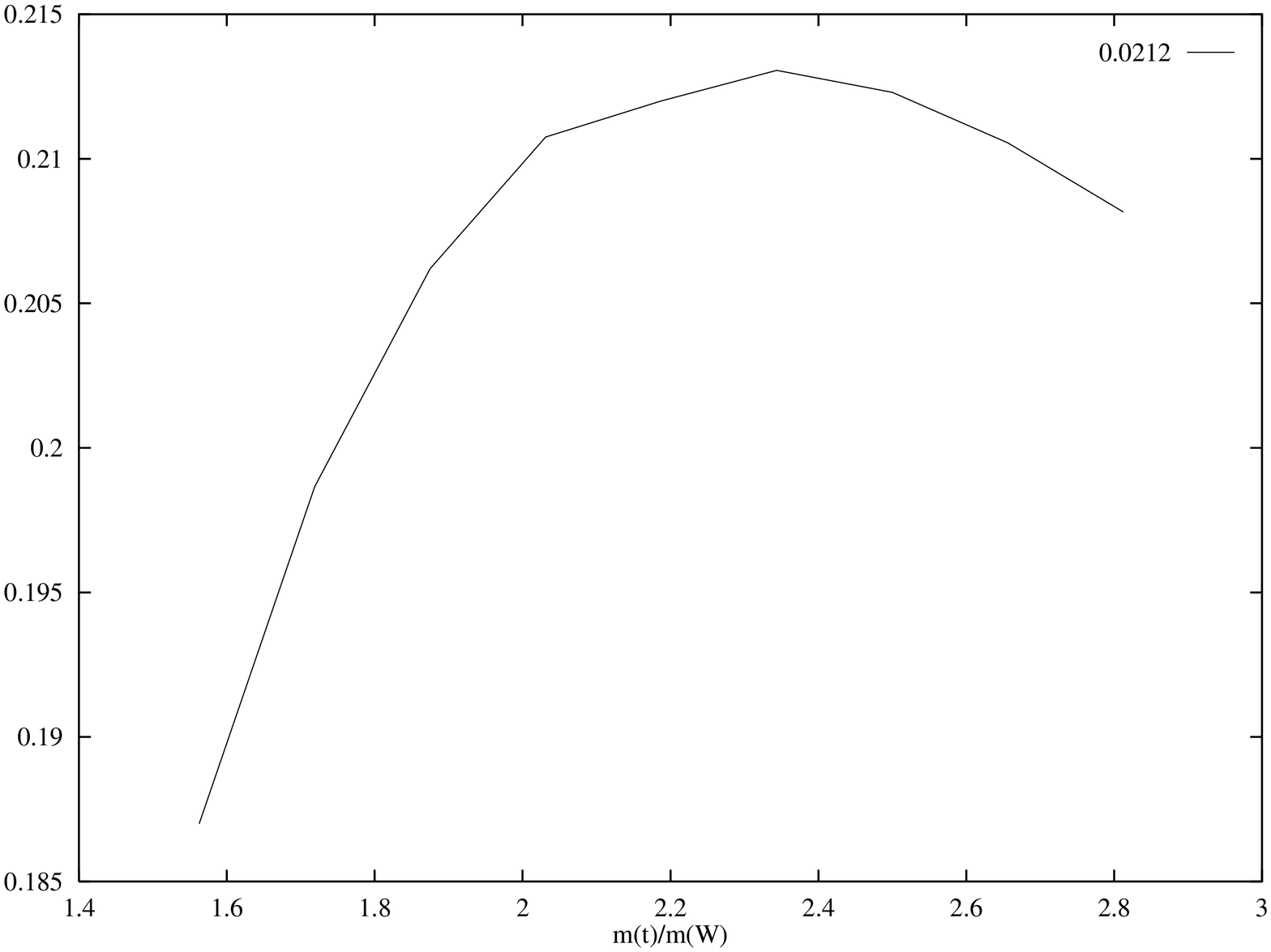,height=10cm,angle=270}
\bigskip
\caption{}
\nonumber
\end{center} 
\end{figure} 

\begin{figure}
\begin{center}  
\epsfig{file=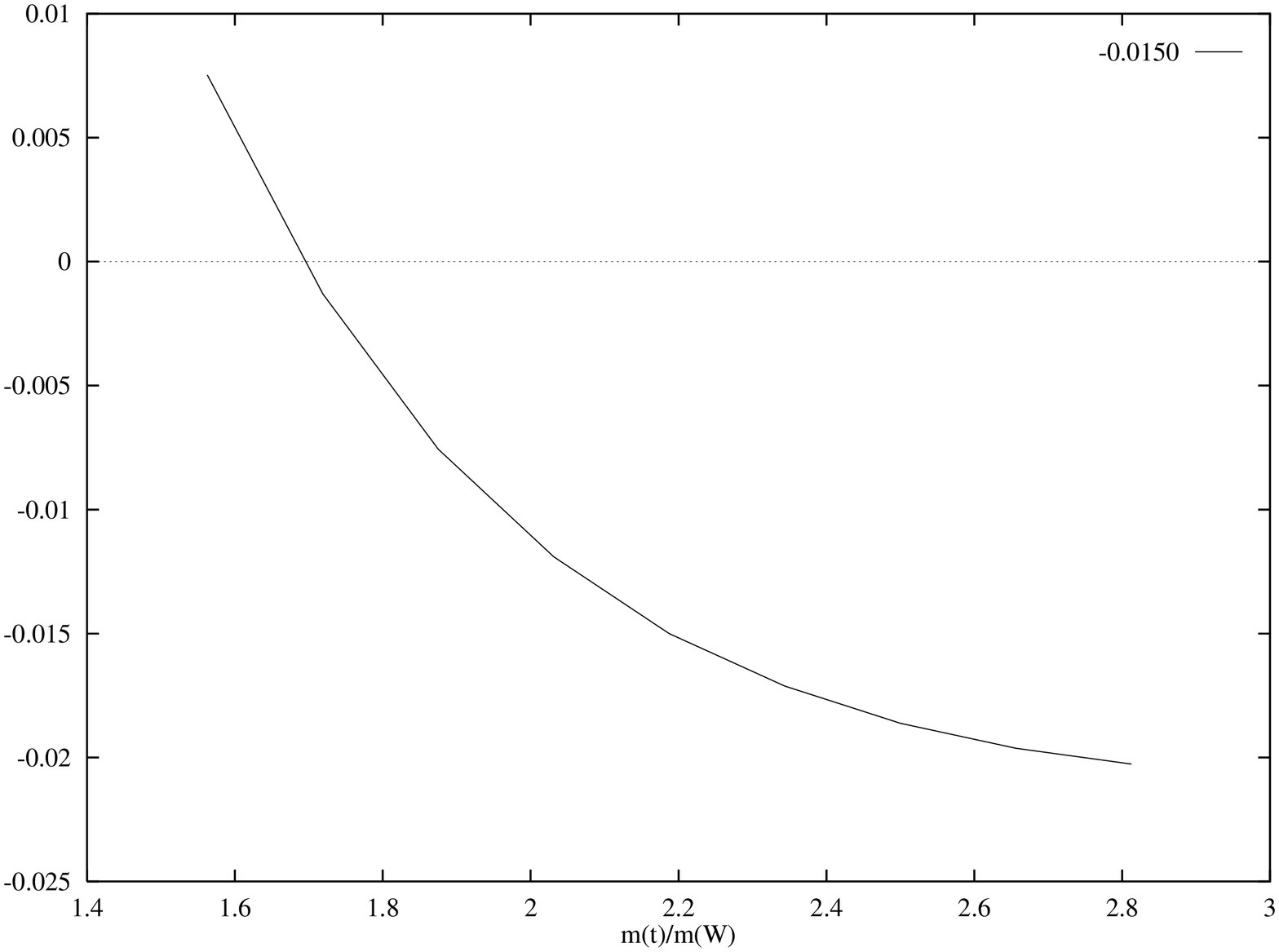,height=10cm,angle=270}
\bigskip
\caption{}
\nonumber
\end{center} 
\end{figure} 

\begin{figure}
\begin{center}  
\epsfig{file=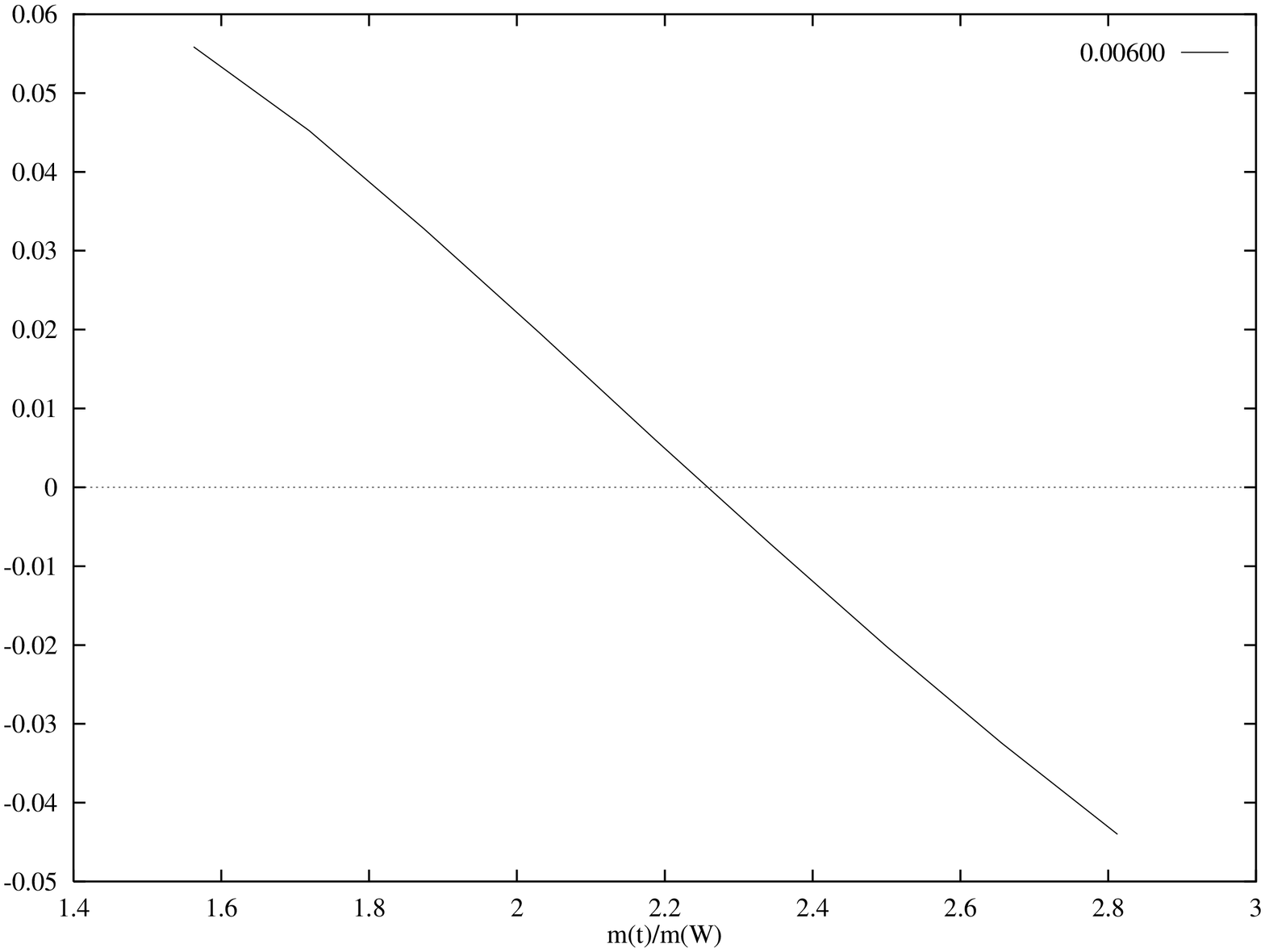,height=10cm,angle=270}
\bigskip
\caption{}
\nonumber
\end{center} 
\end{figure} 

\begin{figure}
\begin{center}  
\epsfig{file=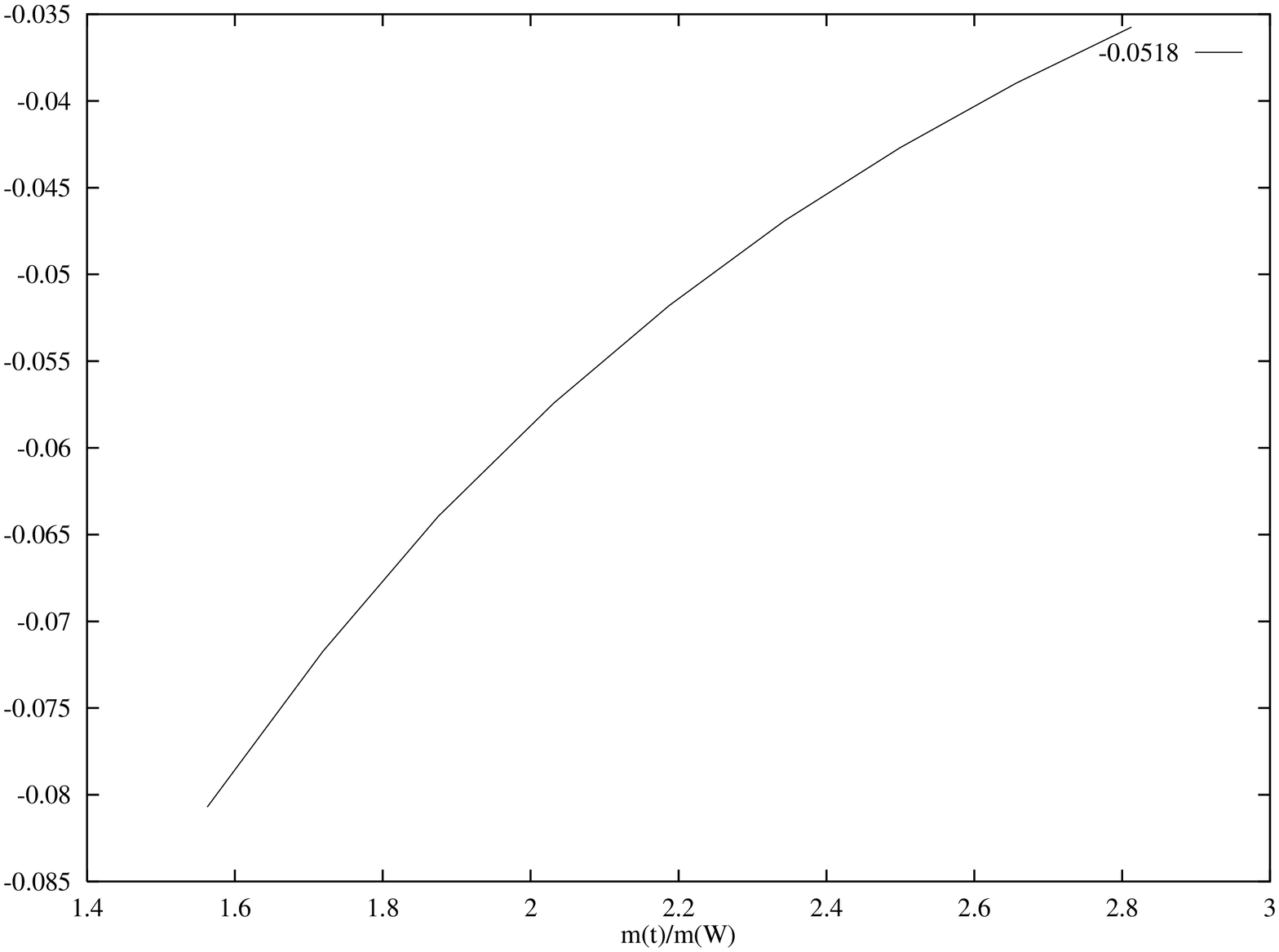,height=10cm,angle=270}
\bigskip
\caption{}
\nonumber
\end{center} 
\end{figure} 

\begin{figure}
\begin{center}  
\epsfig{file=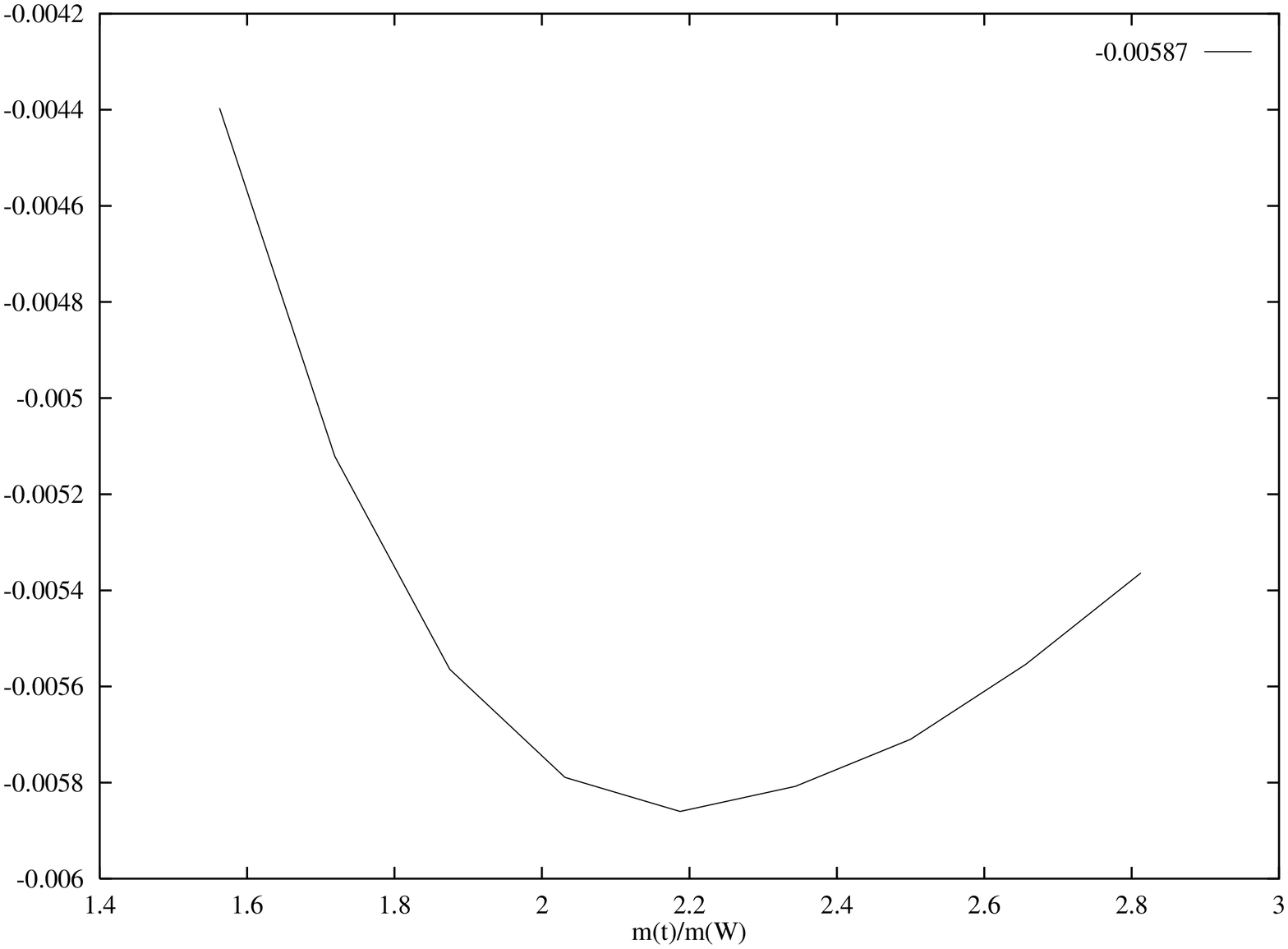,height=10cm,angle=270}
\bigskip
\caption{}
\nonumber
\end{center} 
\end{figure} 

\begin{figure}
\begin{center}  
\epsfig{file=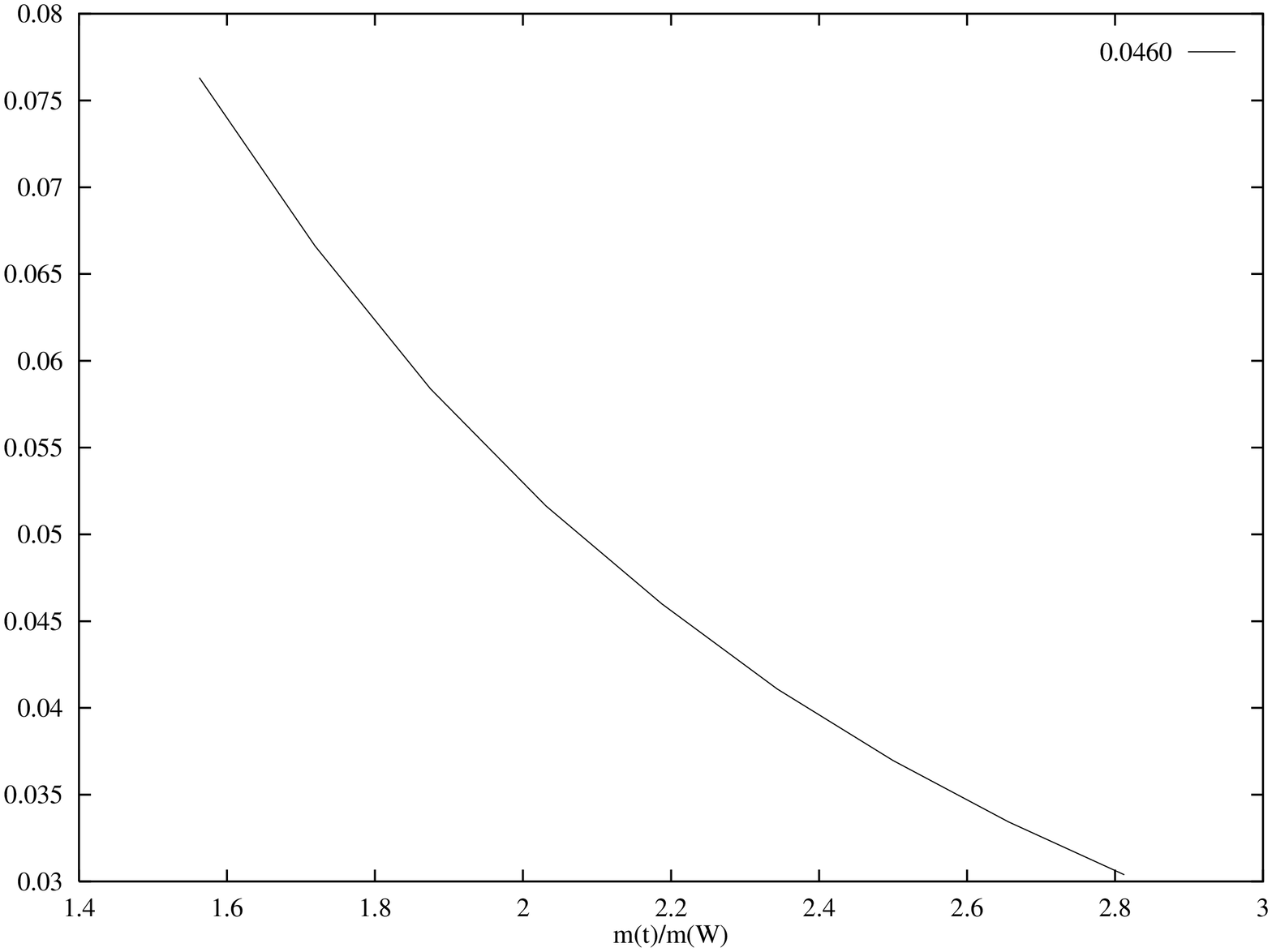,height=10cm,angle=270}
\bigskip
\caption{}
\nonumber
\end{center} 
\end{figure} 

\begin{figure}
\begin{center}  
\epsfig{file=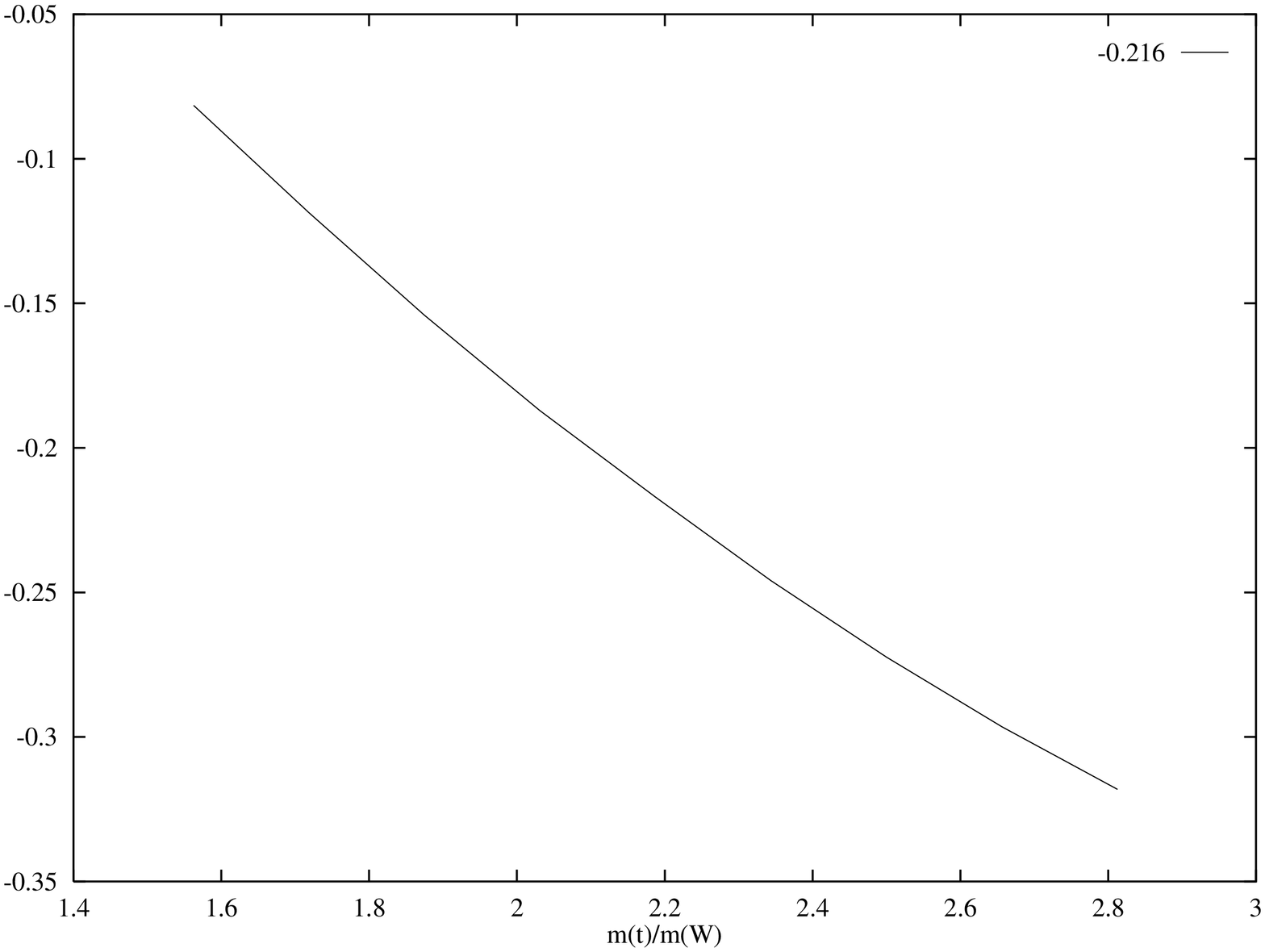,height=10cm,angle=270}
\bigskip
\caption{}
\nonumber
\end{center} 
\end{figure} 

\begin{figure}
\begin{center}  
\epsfig{file=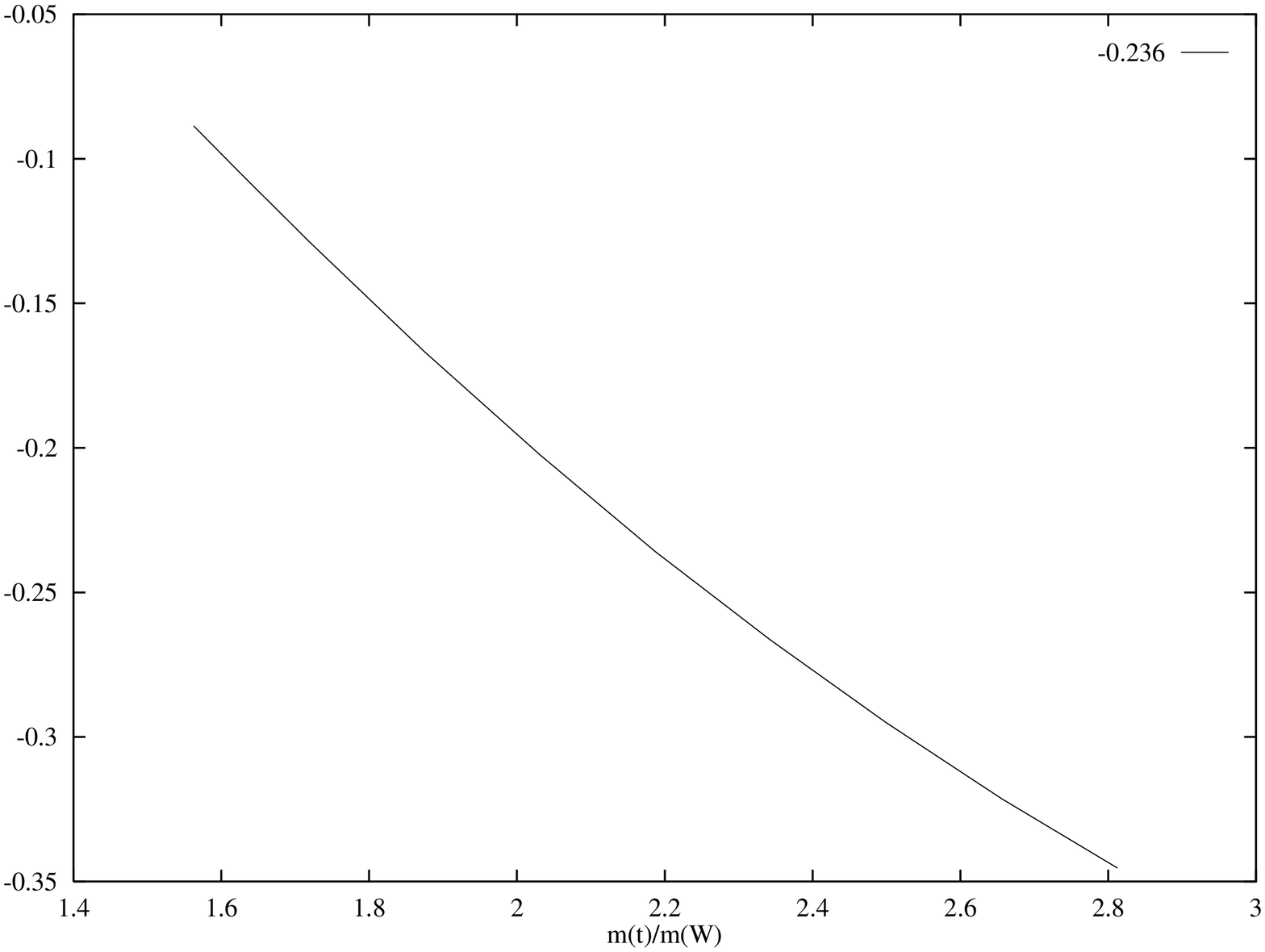,height=10cm,angle=270}
\bigskip
\caption{}
\nonumber
\end{center} 
\end{figure} 

\begin{figure}
\begin{center}  
\epsfig{file=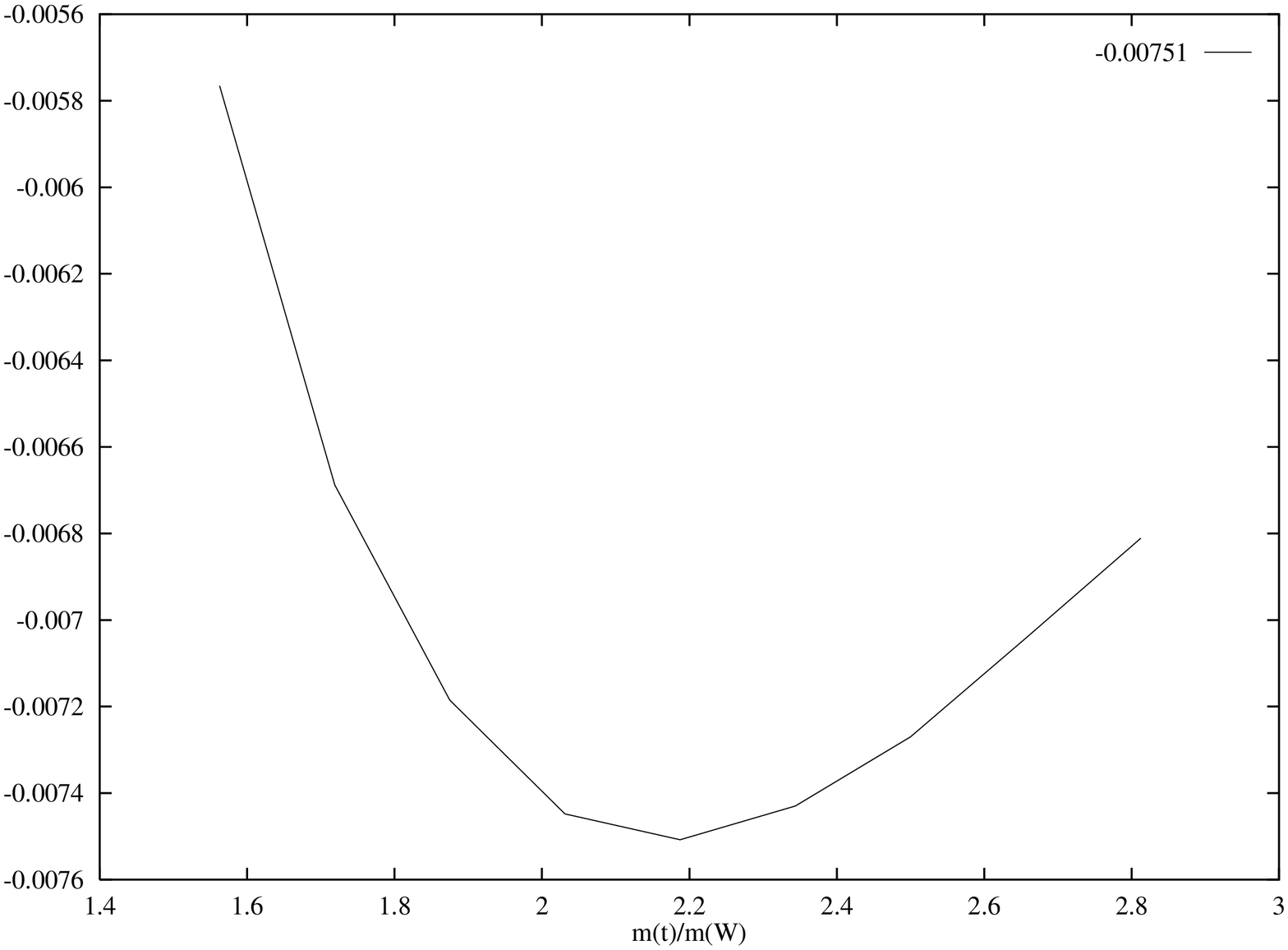,height=10cm,angle=270}
\bigskip
\caption{}
\nonumber
\end{center} 
\end{figure} 

I have carried through a second analogous calculation 
in the rest system of the W and obtained the real gluon QCD corrections 
to the density matrix $\rho_{norm}(h_t,h_W,h_t',h_W')$ also 
in that system. The momenta are now parametrized as follows 
\begin{eqnarray} \nonumber  & & 
p_W=(m_t,\vec 0) \qquad p_t={m_t\over 2\sqrt{f}}f_+x_t(1,0,0,\beta_t)   
\qquad p_g=p_t-p_W-p_b 
\\ & & 
p_b={m_t\over 2\sqrt{f}}f_-x_b
(1,\sin\theta_b\cos\phi_b,\sin\theta_b\sin\phi_b,\cos \theta_b)
\label{real177}
\end{eqnarray}
where 
\begin{equation}   
\beta_t^2=1-{4f\over f_+^2x_t^2}
\label{wd528}       
\end{equation}     
and where $\theta_b$
is given in terms of the other variables according to
\begin{equation}   
f_-x_b+f_+(1-x_t)={1\over 2f}f_+x_tf_-x_b(1-\beta_t\cos \theta_b) \, . 
\label{wd538}       
\end{equation}     
The integrations over the gluon's degrees of freedom are encoded 
as integrations over $x_t$, $x_b$ and $\phi_b$ and have to
be done with the phase space
\begin{equation}
dPS_3(t\rightarrow Wbg)={m_t^2\pi^2\over 32}f_+f_-\int_0^1d(1-x_b)
\int_0^{{f_-^2x_b(1-x_b)\over f_+(1-f_-(1-x_g))}} d(1-x_t)
\int_0^{2\pi} {d\phi_b \over 2\pi}    \, .
\label{wd5477}
\end{equation}
About the effects of gluon emission on polarization: 
Since the W--momentum is unchanged as compared to lowest 
order, the form of the W polarization vectors remains as in   
(\ref{q52}). However, the top and bottom spinors are modified. 
They now read 
\begin{equation}
u_{+1/2}(p_t)=\sqrt{m_t} (a_-,0,a_+,0)
\qquad 
u_{-1/2}(p_t)=\sqrt{m_t} (0,a_-,0,a_+)
\label{q2172}
\end{equation}
with 
\begin{equation}
a_{\pm}=\sqrt{ {x_tf_+\over 2\sqrt{f} } } \sqrt{ 1\pm \beta_t} 
\label{q2171}
\end{equation}
and
\begin{equation}
\bar u_{-1/2}(p_b)=\sqrt{m_t} (0,0,b_1^{\ast},b_2^{\ast})
\label{q2265}
\end{equation}
where $b_1^{\ast}$ and $b_2^{\ast}$ are given indirectly by 
$u_{-1/2}(p_b)\bar u_{-1/2}(p_b)={1\over 2}(1-\gamma_5) b\llap{/}$. 
The 36 elements of the 
normalized density matrix obtained in this frame are given by 
\begin{eqnarray}  & & 
(1+2f)\rho_{norm}(-{1\over2},0,-{1\over2},0)=
\cos^2\theta+(-0.248\, z_- +0.188 \, z_+) \, \alpha_d 
\label{real25a}
\\  & & 
(1+2f)\rho_{norm}(+{1\over2},0,+{1\over2},0)=
f\sin^2\theta+ (-0.080\, z_- -0.0246\, z_+) \, \alpha_d
\\  & & 
(1+2f)\rho_{norm}(-{1\over2},0,+{1\over2},0)=
-\sqrt{f}  \, \sin\theta \cos\theta  \, (1-0.0171  \, \alpha_d ) 
\\  & & 
(1+2f)\rho_{norm}(+{1\over2},0,-{1\over2},0)=
-\sqrt{f}  \, \sin\theta \cos\theta  \, (1-0.0171  \, \alpha_d ) 
%
%
%
%
\\ & & 
(1+2f)\rho_{norm}(-{1\over2},-1,-{1\over2},-1)=
{1\over2}\sin^2\theta  
\\ & & \qquad \qquad \qquad  \qquad 
           +(0.188\, z_- -0.0296\, z_+-0.0222\, \cos\theta ) \, \alpha_d
\label{rek101}
\\ & & 
(1+2f)\rho_{norm}(+{1\over2},-1,+{1\over2},-1)=
{f\over2}(1-\cos\theta)^2
\\ & & \qquad \qquad \qquad  \qquad 
      + (-0.0246\, z_- -0.0522\, z_+ +0.0464\, \cos\theta ) \, \alpha_d
\label{rek102}
\\ & & 
(1+2f)\rho_{norm}(+{1\over2},-1,-{1\over2},-1)=                     
(1+2f)\rho_{norm}(-{1\over2},-1,+{1\over2},-1)                     
\\ & & \qquad    
={\sqrt{f}\over 2} \,  [ \, - \sin\theta  \, (1+0.0640  \, \alpha_d )
                    +\sin\theta \cos\theta  \, (1-0.0374  \, \alpha_d )]
\\  & & 
(1+2f)\rho_{norm}(-{1\over2},+1,-{1\over2},+1)=
{1\over2}\sin^2\theta  
\\ & & \qquad \qquad \qquad  \qquad 
           +(0.188\, z_- -0.0296\, z_++0.0222\, \cos\theta ) \, \alpha_d  
\\ & & 
(1+2f)\rho_{norm}(+{1\over2},+1,+{1\over2},+1)=
{f\over2}(1+\cos\theta)^2
\\ & & \qquad \qquad \qquad  \qquad 
      + (-0.0246\, z_- -0.0522\, z_+ -0.0464\, \cos\theta ) \, \alpha_d
\\ & & 
(1+2f)\rho_{norm}(+{1\over2},+1,-{1\over2},+1)=
(1+2f)\rho_{norm}(-{1\over2},+1,+{1\over2},+1)
\\ & & \qquad   
={\sqrt{f}\over 2}  \, [ \,  \sin\theta  \, (1+0.0293  \, \alpha_d ) 
                    +\sin\theta \cos\theta  \, (1-0.0171  \, \alpha_d )]
\\ & & 
(1+2f)\rho_{norm}(-{1\over2},0,-{1\over2},+1)=
     (1+2f)\rho_{norm}(-{1\over2},+1,-{1\over2},0)
\\ & & \qquad \qquad \qquad 
={1\over \sqrt{2}} \sin\theta \cos\theta (1+0.218  \, \alpha_d )
-0.0157   \,  \alpha_d  \, \sin\theta 
\\ & & 
(1+2f)\rho_{norm}(+{1\over2},0,+{1\over2},+1)=
(1+2f)\rho_{norm}(+{1\over2},+1,+{1\over2},0)
\\ & & \qquad                          
=-{f\over \sqrt{2}}  \, [  \, \sin\theta  \, (1-0.222  \, \alpha_d ) 
                    +\sin\theta \cos\theta  \, (1-0.1325  \, \alpha_d )]
\\ & & 
(1+2f)\rho_{norm}(-{1\over2},0,+{1\over2},+1)=
(1+2f)\rho_{norm}(+{1\over2},+1,-{1\over2},0)
\\ & &                          
=\sqrt{ f\over 2} \cos\theta \, (1+\cos\theta ) 
   +  ( 0.0130  \, z_- +0.00196   \, z_+ -0.00196  \, \cos\theta ) \, \alpha_d
\label{really}
\\ & & 
(1+2f)\rho_{norm}(+{1\over2},0,-{1\over2},+1)=
(1+2f)\rho_{norm}(-{1\over2},+1,+{1\over2},0)
\\ & & \qquad                         
=-\sqrt{ f\over 2} \sin^2\theta 
  +(0.00358 \, z_- -0.00750  \, z_+ +0.00750  \, \cos\theta ) \, \alpha_d 
\\ & & 
(1+2f)\rho_{norm}(-{1\over2},0,-{1\over2},-1)=
(1+2f)\rho_{norm}(-{1\over2},-1,-{1\over2},0)
\\ & & \qquad \qquad \qquad
=-{1\over \sqrt{2}} \sin\theta \cos\theta  \, (1+0.218  \, \alpha_d ) 
-0.0157  \,   \alpha_d  \, \sin\theta 
\\ & & 
(1+2f)\rho_{norm}(+{1\over2},0,+{1\over2},-1)=
(1+2f)\rho_{norm}(+{1\over2},-1,+{1\over2},0)
\\ & & \qquad 
=-{f\over \sqrt{2}}  \, [  \, \sin\theta  \, (1-0.222  \, \alpha_d ) 
                    -\sin\theta \cos\theta  \, (1-0.1325  \, \alpha_d )]
\\ & & 
(1+2f)\rho_{norm}(-{1\over2},0,+{1\over2},-1)=
(1+2f)\rho_{norm}(+{1\over2},-1,-{1\over2},0)
\\ & & 
=\sqrt{ f\over 2} \cos\theta \, (1-\cos\theta )
   +  ( -0.0130  \, z_- -0.00196 \, z_+ +0.00196  \, \cos\theta ) \, \alpha_d
\\ & & 
(1+2f)\rho_{norm}(+{1\over2},0,-{1\over2},-1)=
(1+2f)\rho_{norm}(-{1\over2},-1,+{1\over2},0)
\\ & & \qquad 
=\sqrt{ f\over 2} \sin^2\theta
  +(-0.00358 \, z_- +0.00750  \, z_+ +0.00750  \, \cos\theta ) \, \alpha_d
\\ & & 
(1+2f)\rho_{norm}(-{1\over2},+1,-{1\over2},-1)= 
(1+2f)\rho_{norm}(-{1\over2},-1,-{1\over2},+1)
\\ & & \qquad \qquad \qquad
=-z_-  \, (1+0.218  \, \alpha_d )
\\ & & 
(1+2f)\rho_{norm}(+{1\over2},+1,+{1\over2},-1)=
(1+2f)\rho_{norm}(+{1\over2},-1,+{1\over2},+1)                     
\\ & & \qquad \qquad \qquad
=f \, z_- \, (1-0.1325  \, \alpha_d )
\\ & & 
(1+2f)\rho_{norm}(-{1\over2},+1,+{1\over2},-1)=
(1+2f)\rho_{norm}(+{1\over2},-1,-{1\over2},+1)
\\ & & \qquad \qquad \qquad
={\sqrt{f}\over 2}  \, \sin\theta \,  ( \, 1          
                    - \cos\theta  \, ) \,  (1-0.0171  \, \alpha_d )
\\ & & 
(1+2f)\rho_{norm}(+{1\over2},+1,-{1\over2},-1)=
(1+2f)\rho_{norm}(-{1\over2},-1,+{1\over2},+1)
\\ & & \qquad \qquad \qquad
=-{\sqrt{f}\over 2}  \, \sin\theta \,  ( \, 1                      
                    + \cos\theta  \, ) \, (1-0.0171  \, \alpha_d ) 
\label{real25}              
\end{eqnarray}             
where $z_{\pm}={1\over2}(1\pm \cos^2\theta)$. 
The QCD coefficients have again been obtained by numerical 
integration with $m_t=175$ GeV. 
The $m_t$ dependence of these coefficients is again moderate.
In fact, it can be shown that the coefficients in the rest frame 
of the top quark, Eqs. (\ref{real2a})--(\ref{real2}), 
and of the W boson, Eqs. (\ref{real25a})--(\ref{real25}), are 
related. I was not able to derive a general formula, but I 
have found, for example, that 
$-0.0518$[Eq.(\ref{rek1})]$=-0.0296$[Eq.(\ref{rek101})]$-0.0222$
[Eq.(\ref{rek101})]  and 
$-0.00587$[Eq.(\ref{rek1})]$=-0.0522$[Eq.(\ref{rek102})]$+0.0464$
[Eq.(\ref{rek102})]. 
Corresponding equalities are true for all other values of 
$m_t$, i.e. they hold for the coefficients in general. 
There are some other relations which I do not want to quote here.

\section{Summary}
In this report I have summarized a recent new calculation 
of a complete spin analysis of the Standard Model top quark decay 
including higher order QCD corrections. 
The QCD corrections to the 'normalized' density matrix 
are in general quite small, of the order of 1\%, in particular 
for the real gluon contribution in the rest frame of the top quark, 
cf. Eqs. (\ref{real2a})--(\ref{real2}) and of the W,  
cf. Eqs. (\ref{real25a})--(\ref{real25}). 
The contribution from virtual gluons is somewhat larger, 
cf. Eq. (\ref{vrt1}). 
Note that the {\it relative} magnitude of the QCD corrections 
can be very large -- in all cases, where the lowest order 
contribution vanishes, like Eqs. (\ref{rek1})--(\ref{rek2}). 
In other cases, the symmetry requirement of CP gives vanishing 
matrix elements beyond the leading order, 
cf. Eqs. (\ref{rek3})--(\ref{real2}). 

Complete results including azimuthal 
dependence, 
numerical analysis and 
physical applications 
have not been included here.   
However, I plan 
to write a long article with J. K\"orner and his group 
\cite{new}, in which 
not only this, but also analytical formulae 
for all the QCD coefficients will be given.

\section*{Acknowledgments}
Discussions with Joseph Abraham, J\"urgen K\"orner and Bohdan Grzadskowski 
are gratefully acknowledged.

\end{document}